\documentclass[aps,prd,twocolumn,superscriptaddress,nofootinbib,preprintnumbers]{revtex4-1}
\usepackage{amsmath}
\usepackage{graphicx}
\usepackage{subfigure}
\usepackage{amssymb}
\usepackage{xcolor}
\usepackage{multirow}
\usepackage{cancel}
\usepackage{color}
\usepackage{ulem}
\usepackage{listings}
\usepackage{xcolor}
\usepackage{float}
\usepackage{slashed}

\usepackage{wrapfig}
\usepackage{mathtools}

\usepackage{tikz}
\usetikzlibrary{arrows,shapes}
\usetikzlibrary{trees}
\usetikzlibrary{matrix,arrows} 				
\usetikzlibrary{positioning}				
\usetikzlibrary{calc,through}				
\usetikzlibrary{decorations.pathreplacing}  
\usepackage{pgffor}							

\usetikzlibrary{decorations.pathmorphing}	
\usetikzlibrary{decorations.markings}
\tikzset{
    vector/.style={decorate, decoration={snake}, draw},
	provector/.style={decorate, decoration={snake,amplitude=2.5pt}, draw},
	antivector/.style={decorate, decoration={snake,amplitude=-2.5pt}, draw},
    fermion/.style={draw=black, postaction={decorate},
        decoration={markings,mark=at position .55 with {\arrow[draw=black]{>}}}},
    fermionbar/.style={draw=black, postaction={decorate},
        decoration={markings,mark=at position .55 with {\arrow[draw=black]{<}}}},
    fermionnoarrow/.style={draw=black},
    gluon/.style={decorate, draw=black,
        decoration={coil,amplitude=4pt, segment length=5pt}},
    scalar/.style={dashed,draw=black, postaction={decorate},
        decoration={markings,mark=at position .55 with {\arrow[draw=black]{>}}}},
    scalarbar/.style={dashed,draw=black, postaction={decorate},
        decoration={markings,mark=at position .55 with {\arrow[draw=black]{<}}}},
    scalarnoarrow/.style={dashed,draw=black},
    electron/.style={draw=black, postaction={decorate},
        decoration={markings,mark=at position .55 with {\arrow[draw=black]{>}}}},
	bigvector/.style={decorate, decoration={snake,amplitude=4pt}, draw},
}

\tikzstyle{block} = [draw, rectangle, 
    minimum height=3em, minimum width=6em]

\usepackage[colorlinks,citecolor=blue]{hyperref}
\usepackage{amsmath}
\usepackage{wrapfig}

\usepackage[T1]{fontenc} 
\usepackage[utf8]{inputenc} 
\usepackage{times}

\newcommand{\be}{\begin{equation}}
\newcommand{\ee}{\end{equation}}
\newcommand{\beq}{\begin{equation}}
\newcommand{\eeq}{\end{equation}}
\newcommand{\bea}{\begin{eqnarray}}
\newcommand{\eea}{\end{eqnarray}}
\newcommand{\besp}{\begin{equation}\begin{split}}
\newcommand{\eesp}{\end{split}\end{equation}}

\newcommand{\nn}{\nonumber}

\newcommand{\Eq}[1]{Eq.~(\ref{#1})}

\newcommand{\Dfbd}{\mathord{\buildrel{\lower3pt\hbox{$\scriptscriptstyle\leftrightarrow$}}\over {D}_{\mu}}}
\newcommand{\ave}[1]{\left\langle #1\right\rangle}

\def\mL{\mathcal{L}}

\def\mO{\mathcal{O}}
\def\mP{\mathcal{P}}

\def\mT{\mathcal{T}}

\def\0{\textbf{0}}
\def\1{\textbf{1}}
\def\2{\textbf{2}}
\def\3{\textbf{3}}
\def\4{\textbf{4}}
\def\5{\textbf{5}}
\def\6{\textbf{6}}
\def\7{\textbf{7}}
\def\8{\textbf{8}}
\def\9{\textbf{9}}

\def\d{\text{d}}

\begin{document}

\title{Probing radiative electroweak symmetry breaking with colliders and gravitational waves}

\author{Wei Liu}
\email{wei.liu@njust.edu.cn}
\affiliation{Department of Applied Physics and MIIT Key Laboratory of Semiconductor Microstructure and Quantum Sensing, Nanjing University of Science and Technology, Nanjing 210094, P. R. China}

\author{Ke-Pan Xie}
\email{kpxie@buaa.edu.cn}
\affiliation{School of Physics, Beihang University, Beijing 100191, P. R. China}

\begin{abstract}

Radiative symmetry breaking provides an appealing explanation for electroweak symmetry breaking and addresses the hierarchy problem. We present a comprehensive phenomenological study of this scenario, focusing on its key feature: the logarithmic-shaped potential. This potential gives rise to a relatively light scalar boson that mixes with the Higgs boson and leads to first-order phase transitions (FOPTs) in the early Universe. Our study includes providing exact and analytical solutions for the vacuum structure and scalar interactions, classifying four patterns of cosmic thermal history, and calculating the supercooled FOPT and gravitational waves (GWs). A detailed treatment of the FOPT dynamics reveals that an ultra-supercooled FOPT does not always imply strong GW signals, due to its short duration. By combining future collider and GW experiments, we can probe the conformal symmetry breaking scales up to $10^5-10^8$ GeV.

\end{abstract}

\maketitle

\newpage
\section{Introduction}

The discovery of the Higgs boson at the Large Hadron Collider (LHC)~\cite{ATLAS:2012yve,CMS:2012qbp} represents a significant milestone in understanding fundamental particles and their interactions. However, the mechanism of electroweak symmetry breaking (EWSB) remains a mystery. In the Standard Model (SM), EWSB is achieved through a negative mass squared term in the Higgs potential. While minimal and economical, it lacks a fundamental explanation for this term's origin. This bare mass term is sensitive to ultraviolet (UV) physics, necessitating fine-tuned UV parameters to yield a Higgs mass of 125 GeV. This is the well-known hierarchy problem that has motivated the exploration of physics beyond the SM (BSM), including theories such as supersymmetry, composite Higgs, and extra dimensions.

Radiative symmetry breaking offers a viable explanation for EWSB and addressing the hierarchy problem~\cite{Coleman:1973jx,Jackiw:1974cv,Bardeen:1995kv,Meissner:2007xv,deBoer:2024jne,Frasca:2024fuq}. In this framework, the Lagrangian does not contain dimensionful parameters at tree-level, and hence is classically scale-invariant or conformal.\footnote{Scale transformation is a subset of the entire conformal group. However, scale invariance implies the full conformal invariance in many quantum field theory models~\cite{Nakayama:2013is}. Here we use these two terms interchangeably.} At the one-loop level, radiative corrections induce a logarithmic contribution to the scalar potential, leading to spontaneous symmetry breaking. This effect arises from quantum corrections, characterizing it as an anomaly. It can also be interpreted as dimensional transmutation resulting from the renormalization group running of the scalar couplings.

While the concept of radiative symmetry breaking is appealing, its direct application to the SM without extending the particle content results in a Higgs boson mass of $\lesssim10$ GeV (excluding the top quark contribution) or an unstable electroweak (EW) vacuum (including the top quark contribution), both of which conflict with experimental data. To align with Higgs measurements, the framework has been modified so that radiative symmetry breaking occurs in a BSM sector and is transmitted to the SM via Higgs-portal couplings, thereby inducing EWSB~\cite{Hempfling:1996ht}. This mechanism also presents potential solutions to longstanding problems in particle physics, including neutrino mass~\cite{Iso:2009ss,Iso:2009nw,Chun:2013soa,Das:2015nwk}, matter-antimatter asymmetry~\cite{Khoze:2013oga,Davoudiasl:2014pya,Huang:2022vkf,Chun:2023ezg}, and dark matter~\cite{Okada:2012sg,Hambye:2013dgv,Kang:2020jeg,YaserAyazi:2019caf,Mohamadnejad:2019vzg,Khoze:2022nyt,Frandsen:2022klh,Hambye:2018qjv,Baldes:2018emh,Wong:2023qon} or primordial black holes~\cite{Gouttenoire:2023pxh,Salvio:2023blb,Salvio:2023ynn,Conaci:2024tlc,Banerjee:2024fam}.

In this study, we examine the phenomenology of radiative EWSB. The distinctive feature of this scenario is the logarithmic potential of the BSM scalar field $\phi$, leading to two types of specific phenomenological signals. First, field excitation near the vacuum produces a scalar boson with mass significantly lighter than the BSM scale $w$, which can be detected at current or future particle colliders. Second, the flat potential near the origin results in one or more first-order phase transitions (FOPTs) in the early Universe, creating stochastic gravitational waves (GWs) observable today. By analyzing these signals, we aim to identify the signatures of the radiative symmetry breaking mechanism.

Originating in the 1970s, this mechanism has been extensively explored in collider phenomenology and cosmology. Here we outline the novelties of our research before moving into the details. First, we present the first exact analytical solution for the vacuum structure, scalar mixings, and interactions under the condition that $w\gg$ the EW scale. This enables a new parameterization of the parameter space, using the new scalar mass $m_\phi$ and its mixing angle $\theta$ with the Higgs boson as inputs, and other parameters such as $w$ can be derived. This approach is particularly advantageous for phenomenological studies as it is closely related to experimental observables.

Second, the new $(m_\phi,\theta)$ parameterization allows us to focus on the fundamental interaction between the SM Higgs boson and the new scalar, which is crucial for propagating conformal symmetry breaking into the EW sector. Unlike many studies that assume additional couplings to SM particles (e.g. embedding models in a gauged $U(1)_{B-L}$ or kinetic mixing frameworks such that $Z'$ can couple to SM particles), our approach highlights the key scalar interactions central to radiative EWSB. This focus enables us to examine the core features of this mechanism.

Third, we combine collider and GW searches. On the collider side, the phenomenology is on the Higgs boson and a singlet scalar, which is an extensively studied topic. We interpret existing bounds and projections within the parameter space of radiative EWSB using the $(m_\phi,\theta)$ scheme. We also conduct a parton-level simulation to investigate the singlet at the future 10 TeV muon collider. On the GW side, we perform a detailed analysis of FOPT dynamics, classifying different patterns in the cosmic thermal history and evaluating the associated GWs. An interesting new finding is that extremely strong (ultra-supercooled) FOPTs do not necessarily generate strong GWs, as the short duration of transition can significantly suppress GW production. The $(m_\phi,\theta)$ parameterization offers a comprehensive view of the parameter space, highlighting the complementarity and cross-verification between collider and GW searches.

This article is organized as follows. In Section~\ref{sec:model}, we introduce the benchmark model and analyze its vacuum structure, presenting the $(m_\phi,\theta)$ parameterization scheme, establishing the framework for the phenomenological study. Section~\ref{sec:collider} focuses on collider phenomenology, while Section~\ref{sec:GW} investigates the dynamics of FOPT and the generation of GWs. We combine the findings from the collider and GW analyses, presenting the final results in Section~\ref{sec:results}. The conclusion is given in Section~\ref{sec:conclusion}.

\section{The model}\label{sec:model}

The tree-level joint potential of the SM Higgs doublet $H=\left(G^+,(h+iG^0)/\sqrt{2}\right)^T$ and the real scalar field $\phi$ reads
\be\label{V0}
V_0(H,S)=\lambda_h|H|^4+\frac{\lambda_s}{4}\phi^4+\frac{\lambda_{h\phi}}{2}|H|^2\phi^2,
\ee
where all the coefficients are dimensionless, and hence the theory is classically conformal. One-loop correction generates logarithmic contributions to \Eq{V0}, known as the Coleman-Weinberg potentials~\cite{Coleman:1973jx,Jackiw:1974cv,Gildener:1976ih}. In principle, both $H$ and $\phi$ receive radiative corrections, resulting in a complicated joint potential. However, under the assumption that the BSM scale significantly exceeds the EW scale and that the magnitude of Higgs-portal coupling $|\lambda_{h\phi}| \ll 1$, we can establish a sequential symmetry breaking scenario~\cite{Chataignier:2018kay}: the BSM radiative correction to the $\phi$-direction generates the spontaneous symmetry breaking at a high scale, which then induces a tree-level potential along the $h$-direction via the $\lambda_{h\phi}$-term, producing the EWSB.

In this case, the potential in unitary gauge up to one-loop level can be written as
\be\label{CC_V}
V_1(h,\phi)=\frac{B}{4}\phi^4\left(\log\frac{\phi}{w_0}-\frac14\right)+\frac{\lambda_{h\phi}}{4}h^2\phi^2+\frac{\lambda_h}{4}h^4,
\ee
which implies different dominance of tree- and loop-level contributions in different field directions. Along the $\phi$-direction, the loop-induced potential dominates and generates the conformal symmetry breaking, resulting in $\ave{\phi}\approx w_0$ and a scalar boson with a mass of $m_\phi\approx\sqrt{B}w_0$. While along the $h$-direction, it is the tree-level contribution that dominates: after $\phi$ acquires its vacuum expectation value (VEV), a Mexican-hat-shaped Higgs potential
\be\label{approxV}
V_h(h)\approx V_1(h,w_0)
=\frac14(\lambda_{h\phi}w_0^2h^2+\lambda_hh^4)
\ee
is generated. Setting the parameters as $\lambda_{h\phi}\approx-m_h^2/w_0^2$ and $\lambda_h\approx m_h^2/(2v^2)$ with $m_h=125$ GeV and $v=246$ GeV, we then get the EWSB with correct Higgs mass and VEV.

The parameter $B$ is contributed by the new physics degrees of freedom in the BSM sector. In the minimal model-building sense, there are two alternative scenarios: gauge-induced or scalar-induced. In the former case, $\phi$ is embedded to a complex scalar $S=(\phi+i\eta)/\sqrt{2}$ that is charged under a gauged dark $U(1)_X$ with the coupling constant $g_X$; while in the latter case, $\phi$ couples to a dark real scalar $X$ via the quartic interaction $\lambda_X\phi^2X^2/4$ in the tree-level potential. Then
\be\label{B}
B=\begin{dcases}~\frac{3g_X^4}{8\pi^2},&\text{for the gauge-induced scenario;}\\
~\frac{\lambda_X^2}{32\pi^2},&\text{for the scalar-induced scenario.}
\end{dcases}
\ee
After the symmetry breaking, the $U(1)_X$ gauge boson $Z'$ or dark scalar $X$ gets a mass of $m_{Z'}\approx g_Xw_0$ or $m_X\approx\sqrt{\lambda_X}w_0/\sqrt{2}$, respectively. BSM fermions coupling to $\phi$ make negative and suppressed contributions to $B$. For example, if we embed the model into a gauged $U(1)_{B-L}$ framework in which $g_X=2g_{B-L}$ and the right-handed neutrino interactions read $-\sum_i y_iS\bar\nu_{R}^{i,c}\nu_{R}^i/2+{\rm h.c.}$~\cite{Iso:2009ss,Iso:2009nw}, then $B\propto(g_{B-L}^4-\sum_iy_i^4/96)$. Therefore, the bosonic BSM degrees of freedom dominate $B$, and we will consider the two minimal realizations in \Eq{B} as research benchmarks. As will be demonstrated, the particle phenomenology of $\phi$ is independent of the source of $B$, and the GW signals are likewise insensitive to its origin. Therefore, we will not specify the explicit expression for $B$ in our discussion unless necessary.

While the above description shows a very clear qualitative picture of the symmetry breaking pattern, it neglects the impact of the Higgs-portal coupling on the $\phi$-direction potential, which causes the mixing between $\phi$ and $h$. That is why we used ``$\approx$'' when discussing the VEVs and particle masses around \Eq{approxV}. Below we resolve the vacuum structure using the full expression of \Eq{CC_V}, providing the exact and analytical solution for scalar VEVs, masses, and mixing angle. Let $\ave{h}=v$ and $\ave{\phi}=w$ be the vacuum where $\partial V_1/\partial h$ and $\partial V_1/\partial\phi$ vanish, we find
\be\label{lambdaph_CC}
\lambda_{h\phi}=-2B\frac{w^2}{v^2}\ln\frac{w}{w_0},\quad \lambda_h=B\frac{w^4}{v^4}\ln\frac{w}{w_0}.
\ee
Since $\lambda_h>0$ is required from the bounded below condition, one can infer $w>w_0$, thus the $\phi$ VEV is larger than the bare parameter $w_0$.

Next, we diagonalize the Hessian matrix
\bea\label{hes}
{\rm Hes}&=&\begin{pmatrix}\frac{\partial^2V_1}{\partial h^2} \,& \frac{\partial^2V_1}{\partial h\partial\phi}\\[0.3cm] \frac{\partial^2V_1}{\partial h\partial\phi} \,& \frac{\partial^2V_1}{\partial\phi^2}\end{pmatrix}_{(v,w)}\\
&=&
\begin{pmatrix}
3\lambda_hv^2+\frac{\lambda_{h\phi}}{2}w^2 & \lambda_{h\phi} vw \\[0,3cm] \lambda_{h\phi} vw & Bw^2+3Bw^2\ln\frac{w}{w_0}+\frac{\lambda_{h\phi}}{2}v^2
\end{pmatrix},\nn
\eea
to get the two scalar mass eigenvalues
\begin{multline}\label{mh1mh2_CC}
m_{h_{1,2}}^2=\frac{w^2}{4} \left(6 B \ln\frac{w}{w_0}+2 B+\lambda_{h\phi}\right)\\ +\frac{v^2}{4} (6 \lambda_h+\lambda_{h\phi})\mp\frac{\Xi^{1/2}}{4},
\end{multline}
where
\begin{multline}
\Xi=\left[w^2 (2 B-\lambda_{h\phi})+6
B w^2 \ln\frac{w}{w_0}-v^2 (6 \lambda_h-\lambda_{h\phi})\right]^2\\
+16\lambda_{h\phi}^2 v^2 w^2.
\end{multline}
Note $m_{h_2}>m_{h_1}$ by definition, but we do not specify which one is the SM Higgs boson yet.

Substituting \Eq{lambdaph_CC} into \Eq{mh1mh2_CC} to cancel $B$, one obtains
\be\label{xi}
\left(1+\frac{w^2}{v^2}\right)^2\xi^2+\left(1-\frac{(m_{h_1}^4+m_{h_2}^4)w^2}{2m_{h_1}^2m_{h_2}^2v^2}\right)\xi+\frac14=0,
\ee
where $\xi=\ln(w/w_0)$. Resolving this univariate quadratic equation, one obtains two solutions
\bea\label{logww0}
\ln\frac{w}{w_0}&=&\frac{m_{h_1}^4 v^2 w^2+m_{h_2}^4 v^2 w^2-2 m_{h_1}^2 m_{h_2}^2 v^4
\pm\Delta}{4m_{h_1}^2 m_{h_2}^2 \left(v^2+w^2\right)^2},\nn \\
\Delta&=&v^2 w^2\left(m_{h_2}^4-m_{h_1}^4\right) \sqrt{1-\frac{4m_{h_1}^2 m_{h_2}^2v^2}{(m_{h_2}^2-m_{h_1}^2)^2 w^2}}.\qquad
\eea
They correspond to two branches of the physical cases: the ``$+$'' branch is for a singlet lighter than Higgs while the ``$-$'' branch is for a singlet heavier than Higgs. Also note that the definition of $\Delta$ requires
\be\label{non-degenerate}
m_{h_2}>m_{h_1}\left(\sqrt{1+\frac{v^2}{w^2}}+\frac{v}{w}\right),
\ee
thus the two scalar bosons cannot be degenerate.

One of $h_1$ and $h_2$ corresponds to the Higgs boson observed at the LHC, while the other represents the new singlet-like boson yet to be discovered. For simplicity, we use the notations $h$ and $\phi$ to denote the Higgs and singlet-like mass eigenstates, respectively, and define the rotation matrix as
\be
\begin{pmatrix}h\\ \phi\end{pmatrix}\xrightarrow[\rm eigenstates]{\rm to~mass} U\begin{pmatrix}h\\ \phi\end{pmatrix},\quad U=\begin{pmatrix}\cos\theta&\sin\theta\\ -\sin\theta&\cos\theta\end{pmatrix},
\ee
where $\theta$ is the mixing angle satisfying $|\theta|<\pi/4$ so that the magnitude of the diagonal elements is larger than that of the non-diagonal ones. The Hessian matrix is diagonalized as $U^\dagger{\rm Hes}\,U={\rm diag}\{m_h^2,m_\phi^2\}$, with $m_\phi$ being a free parameter that can be either larger or smaller than $m_h$. The two branches of \Eq{logww0} can be summarized as
\bea\label{param}
\lambda_{h\phi}&=&-\frac{w (m_\phi^2+m_h^2)\pm\sqrt{w^2 (m_h^2-m_\phi^2)^2-4 m_\phi^2 m_h^2 v^2}}{2 w \left(v^2+w^2\right)},\nn\\
B&=&\frac{m_\phi^2+m_h^2}{2w^2}\mp\sqrt{\frac{(m_h^2-m_\phi^2)^2}{4w^4}-\frac{m_\phi^2 m_h^2 v^2}{w^6}},\\
\tan\theta&=&\frac{\pm2m_h^2v}{\pm(m_h^2-m_\phi^2)w+\sqrt{w^2(m_h^2-m_\phi^2)^2-4 m_\phi^2 m_h^2 v^2}},\nn
\eea
and $\lambda_h=-\lambda_{h\phi}w^2/(2v^2)$, where the upper sign is for $m_\phi<m_h$, and the lower sign is for $m_\phi>m_h$.

So far, we have changed the input of \Eq{CC_V} from bare parameters $\{B,w_0,\lambda_{h\phi},\lambda_h\}$ to physical observables $\{v,w,m_h,m_\phi\}$, leaving $w$ and $m_\phi$ as the only two free parameters.\footnote{The radiative EWSB model considered here fundamentally differs from the conventional singlet extension of the SM (xSM), which utilizes a polynomial potential with bare mass terms at tree-level~\cite{Lewis:2017dme}. Unlike the xSM, this model does not suffer from the hierarchy problem, and the scalar potential in \Eq{CC_V} implies additional new physics degrees of freedom that contribute to $B$, such as $Z'$ or $X$, while xSM does not necessarily include those particles. Notably, the minimal radiative EWSB model requires only two free parameters, fewer than the five and three required in the real-singlet~\cite{Liu:2021jyc} and complex-singlet~\cite{Li:2023bxy} extensions of the SM, respectively.} When $w\gg v$, expressions become independent of the mass hierarchy between $h$ and $\phi$. For instance, the portal coupling $\lambda_{h\phi}\approx-m_h^2/w^2$, matches our previous estimates; the mixing angle $\tan\theta\approx(v/w)m_h^2/(m_h^2-m_\phi^2)$, consistent with the approximate result in the literature~\cite{Iso:2009nw}. For the convenience of the phenomenological study, we will use $m_\phi$ and $\theta$ as input parameters hereafter, and all other parameters can be derived. For example, the $\phi$ VEV is
\be
w=\frac{m_h^2\cot\theta+m_\phi^2\tan\theta}{|m_h^2-m_\phi^2|}v,
\ee
implying $m_\phi\neq m_h$, consistent with \Eq{non-degenerate}.

\section{Collider phenomenology}\label{sec:collider}

In the minimal setup, the radiative EWSB scenario only contains a new scalar $\phi$ and a possible boson responsible for the Coleman-Weinberg potential in the BSM sector, such as $Z'$ or $X$. This distinguishes its phenomenology from other models addressing the hierarchy problem, like supersymmetry or composite Higgs, which typically predict multiple new physics particles -- superpartners or composite resonances -- at the TeV scale. Additionally, the mass ratios are $m_\phi/m_{Z'}\approx\sqrt{6}g_X/(4\pi)$ and $m_\phi/m_X\approx\sqrt{\lambda_X}/(4\pi\sqrt{2})$, indicating $\phi$ is much lighter than other BSM particles in the perturbative regime where $g_X^2$, $\lambda_X \ll 4\pi$. As a consequence, the expected collider signals at the TeV scale involve a new scalar that mixes with the Higgs boson.

The BSM sector may have other interactions with the SM particles, resulting in additional signals. For instance, if we identify the $U(1)_X$ as $U(1)_{B-L}$, then $Z'$ couples to the quarks and leptons~\cite{Iso:2009ss,Iso:2009nw,Chun:2013soa,Das:2015nwk}. In this research, we focus on the core idea of the radiative EWSB without adding additional BSM interactions, except the assumption that $Z'$ or $X$ can decay into SM or BSM particles, thereby excluding them as dark matter candidates (otherwise additional constraints are imposed on the parameter space and only one free parameter is left). Therefore, our main text only investigates the interactions between the $\phi$ and $h$ bosons derived using the results in Section~\ref{sec:model}. For example, when $v\ll w$ the triple interactions are
\begin{multline}\label{L3}
\mL_3\approx-\frac{m_h^2}{2v}h^3+\frac{m_h^2(2m_h^2+ m_\phi^2)}{2w( m_\phi^2- m_h^2)}h^2 \phi
\\+\frac{m_h^2 m_\phi^2 v(m_h^2-4 m_\phi^2)}{2 w^2 (m_h^2-m_\phi^2)^2}h \phi^2-\frac{5 m_\phi^2}{6 w}\phi^3,
\end{multline}
from which the Feynman rules can be directly read.

The lightest BSM particle $\phi$ couples to the SM fermions and gauge bosons via the mixing with the Higgs boson. Consequently, it decays to SM particles, and the branching ratios depend solely on $m_\phi$ when $m_\phi<2m_h$, which are already well-known in the literature~\cite{Gershtein:2020mwi,Djouadi:2005gi}: the dominant decay channels in different mass ranges are listed as
\be\label{brs}
\begin{dcases}~\text{$e^+e^-$ or $\mu^+\mu^-$},&\text{$m_\phi\lesssim2m_\pi$};\\
~\text{Mesons or $gg$},&\text{$2m_\pi\lesssim m_\phi\lesssim2m_\tau$};\\
~\text{$\tau^+\tau^-$,}&\text{$2m_\tau\lesssim m_\phi\lesssim2m_b$};\\
~\text{$b\bar b$,}&\text{$2m_b\lesssim m_\phi\lesssim2m_W$};\\
~\text{$VV$, with $V=W^\pm$ or $Z$,}&\text{$2m_W\lesssim m_\phi\lesssim2m_h$.}\\
\end{dcases}
\ee
For $m_\phi>2m_h$, the $\phi\to hh$ decay should be included and the partial width is
\be
\Gamma(\phi\to hh)\approx\frac{\mu_{\phi hh}^2}{8\pi m_\phi}\sqrt{1-\frac{4m_h^2}{m_\phi^2}},
\ee
where $\mu_{\phi hh}$ is the coefficient of the $h^2\phi$ term in \Eq{L3}. The $hh$ and $VV$ channels dominate the $m_\phi>2m_h$ region, while the $h\to t\bar t$ channel should be included if $m_\phi>2m_t$.

The search strategy for $\phi$ varies across different mass regions. For a light $\phi$ with $m_\phi\lesssim2m_\tau$, existing bounds have constrained the magnitude of mixing angle $|\theta|$ to be so small that the total decay width $\propto \sin^2\theta$ is tiny, making $\phi$'s lifetime significantly long. Consequently, the long-lived particle (LLP) search can effectively probe this parameter region, with numerous measurements and proposed studies available~\cite{Batell:2022dpx}. For $m_\phi\gtrsim2m_\tau$, we consider detecting $\phi$ via prompt decay at the LHC or a future 10 TeV muon collider. At the LHC, $\phi$ is primarily produced via gluon gluon fusion, with the cross section expressed as $\sigma_h(m_\phi)\times\sin^2\theta$, where $\sigma_h(m_\phi)$ is the production cross section of a SM Higgs with mass at $m_\phi$~\cite{Anastasiou:2016hlm}. The most stringent bound is the $\phi\to ZZ$ search by the CMS collaboration at $\sqrt{s}=13$ TeV with an integrated luminosity of 35.9 fb$^{-1}$~\cite{CMS:2018amk}, which we rescale to 3000 fb$^{-1}$ to make the HL-LHC projection.

Recent research on multi-TeV muon colliders highlights their potential to combine the advantages of hadron and lepton colliders, offering both high collision energies and low backgrounds~\cite{Delahaye:2019omf, Aime:2022flm, Accettura:2023ked}. At multi-TeV muon colliders, the cross section for the vector boson fusion (VBF) process
\be
\mu^+\mu^-\to\begin{dcases}~\phi\nu_\mu\bar\nu_\mu,&\text{$W^+W^-$ fusion;}\\
~\phi\mu^+\mu^-,&\text{$ZZ$ fusion,}
\end{dcases}
\ee
is significantly larger than that for the associated production $\mu^+\mu^-\to Z\phi$~\cite{Han:2020uid}, and hence we consider VBF as the primary production channel for $\phi$. We implement the model with {\tt FeynRules}~\cite{Alloul:2013bka} and output the model file to generate parton-level events using {\tt MadGraph5\_aMC@NLO}~\cite{Alwall:2014hca}. Based on \Eq{brs}, we study $\phi\to\tau^+\tau^-$, $b\bar b$, and $VV$ decay channels for various mass ranges of $\phi$, focusing on fully hadronic final states. A conservative 10\% smearing is applied to the quark, gluon, and tau momenta to mimic jets.

\begin{figure}
\begin{center}
\includegraphics[scale=0.475]{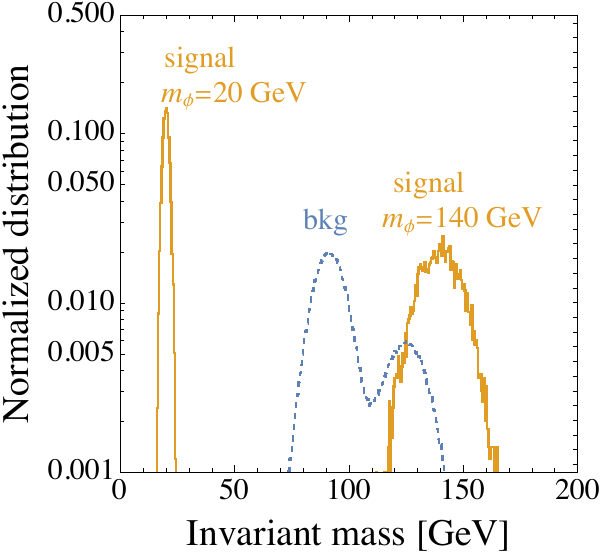}
\caption{Invariant mass of the di-jet system after the basic cuts for the $\phi\to b\bar b$ channel. The blue curves represent the SM VBF $jj$ backgrounds, while the orange curves are the signal distributions for $m_\phi=20$ and 140 GeV.}
\label{fig:jj}
\end{center}
\end{figure}

The VBF $\phi\to \tau^+\tau^-$ and $b\bar b$ channels share the same main background: the SM VBF production of di-jet from photon splitting or $V$ and $h$ decays. We require both the signal and background events to have exactly two jets and no charged leptons with transverse momentum and pseudo-rapidity
\be\label{basic}
p_T>30~{\rm GeV},\quad |\eta|<2.43,
\ee
and recoil mass
\be
m_{\rm recoil}^{jj}=\sqrt{\left(p_{\mu^+}+p_{\mu^-}-p_{j_1}-p_{j_2}\right)^2}>200~{\rm GeV}.
\ee
The cut on $\eta$ corresponds to a detector angle coverage of $[10^\circ,170^\circ]$. The event distributions of the di-jet invariant mass $m_{jj}$ in the $\phi\to b\bar b$ channel after basic cuts are displayed in Fig.~\ref{fig:jj}, where the blue curve clearly shows the peaks of $m_{V,h}$ of the background, while the two orange curves demonstrate the signal peaks for $m_{\phi}=20$ and 140 GeV. A mass-shell cut
\be
|m_{jj}-m_\phi|<{\rm min}\{0.2m_\phi,~30~{\rm GeV}\},
\ee
can efficiently select signal agains the backgrounds, as illustrated in the cut flow of Table~\ref{tab:jj}. We do not assume $b$-tagging in this simulation, but have checked that a 70\% tagging rate yields similar results. For the $\tau^+\tau^-$ channel, we have included the tau hadronic decay branching ratio $\sim65\%$ and assumed a $90\%$ tagging rate.

\begin{table}
\small\centering\renewcommand\arraystretch{1.2}
\begin{tabular}{c|ccc}\hline
Cross sections [fb] & $\sigma_S^{20}$ & $\sigma^{140}_S$ & $\sigma_{B}$ \\ \hline
No Cut & 8.64 & 4.58 & 2870 \\
Basic cuts & 2.87 & 2.34 & 1366 \\
Mass-shell cut, 20  & 2.85 &   & 0.207 \\
Mass-shell cut, 140  &   & 2.33 &  343 \\ \hline
\end{tabular}
\caption{Cut flows for the $\phi\to b\bar b$ channel with $m_\phi=$ 20, 140 GeV and the backgrounds. For the signals, we assume $\sin\theta =0.1$.}
\label{tab:jj}
\end{table}

\begin{figure}
\begin{center}
\includegraphics[scale=0.475]{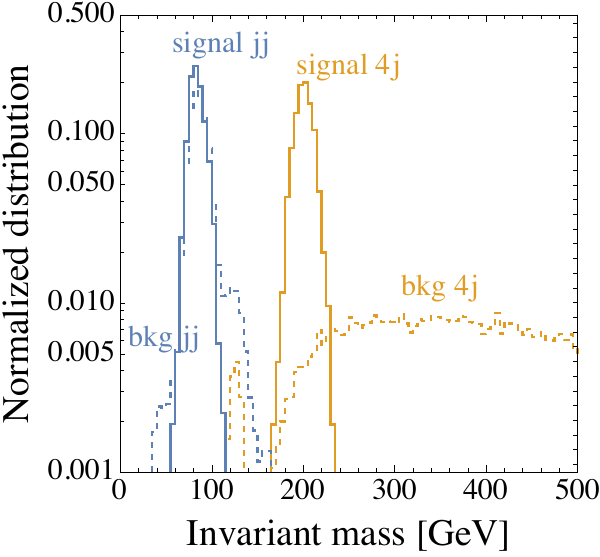}
\caption{Invariant mass of the $jj$ (blue) and $4j$ (orange) systems after the basic cuts for the $\phi\to VV\to jjjj$ channel. The dashed curves represent the SM VBF $jjjj$ backgrounds, while the solid curves are the signal distributions for $m_\phi=200$ GeV.}
\label{fig:vv}
\end{center}
\end{figure}

In the VBF $\phi\to VV\to jjjj$ channel, the main background is the SM VBF $jjjj$ from pure EW process or involving QCD gluon splitting. We apply the following basic cuts: exactly four jets and no charged leptons within the kinetic region of \Eq{basic}, and the recoil mass
\be
m_{\rm recoil}^{4j}=\sqrt{\left(p_{\mu^+}+p_{\mu^-}-\sum_{n=1}^4p_{j_n}\right)^2}>200~{\rm GeV}.
\ee
Then we pair the four jets by minimizing
\begin{multline}
\chi^2={\rm min}\left\{\frac{(m_{j_1j_2}-m_W)^2}{\Gamma_W^2}+\frac{(m_{j_3j_4}-m_W)^2}{\Gamma_W^2},\right.\\
\left.~\frac{(m_{j_1j_2}-m_Z)^2}{\Gamma_Z^2}+\frac{(m_{j_3j_4}-m_Z)^2}{\Gamma_Z^2}\right\},
\end{multline}
where $\Gamma_{W,Z}$ are respectively the decay widths of the $W^\pm$ or $Z$ bosons. After pairing, $(j_1j_2)$ and $(j_3j_4)$ are identified as two $V$ candidates, as illustrated in blue peaked curves of Fig.~\ref{fig:vv}. Note that the main background, the SM EW VBF production of $jjjj$, also has a peak at $\sim m_V$. However, the invariant mass of the entire $4j$ system peaks at $m_\phi$ for the signal, while the background has a mainly smooth distribution plus a small peak at $\sim m_h$ from the $h\to WW^*/ZZ^*\to jjjj$ decay, as shown in the orange curves. Therefore, the mass-shell cuts
\be
|m_{j_1j_2}-m_V|<15~{\rm GeV},\quad
|m_{j_3,j_4}-m_V|<15~{\rm GeV}
\ee
for the $V$ candidates and
\be
|m_{4j}-m_\phi|<30~{\rm GeV}
\ee
for the $4j$ system can efficiently remove background events and manifest the signal, as illustrated in Table~\ref{tab:vv}.

\begin{table}
\small\centering\renewcommand\arraystretch{1.2}
\begin{tabular}{c|cccc}\hline
Cross sections [fb] & $\sigma_S^{200}$ & $\sigma_{B}^{\rm EW}$ & $\sigma_{B}^{\rm QCD}$ \\ \hline
No Cut & 3.09 & 157 & 26.7 \\
Basic cuts & 0.481 & 39.7 & 5.41 \\
Mass-shell cut for $V$  & 0.395 & 23.8 & 0.0615 \\ 
Mass-shell cut for $\phi$ & 0.394 & 1.69 & 0.0246 \\ \hline
\end{tabular}
\caption{Cut flows for the $\phi\to VV$ channel with $m_\phi=200$ GeV and the backgrounds. For the signal, we assume $\sin\theta =0.1$.}
\label{tab:vv}
\end{table}

For $m_\phi>2m_h$, the $\phi\to hh\to b\bar bb\bar b$ channel is the most effective probe of the model, with the main background being the SM VBF $jjjj$. We utilize the simulation results from Ref.~\cite{Liu:2021jyc}, which is based on the xSM and is not for classically conformal models; however, the technical considerations are the same with the radiative EWSB model for this specific channel.

\section{Cosmological implications}\label{sec:GW}

While particle experiments effectively probe the excitations of the quantum field near the vacuum, the signals cannot be considered definitive evidence for the radiative EWSB mechanism. The phenomenology discussed in Section~\ref{sec:collider} primarily addresses a new scalar boson mixing with the Higgs boson, a prediction shared by many new physics models. In this section, we focus on the distinctive feature of the radiative EWSB mechanism: the logarithmic shape of the $\phi$-direction potential. We will demonstrate this shape leads to one or more FOPTs during cosmic evolution, resulting in GW signals.

\subsection{Thermal history}

\begin{figure*}
\begin{center}
\includegraphics[scale=0.34]{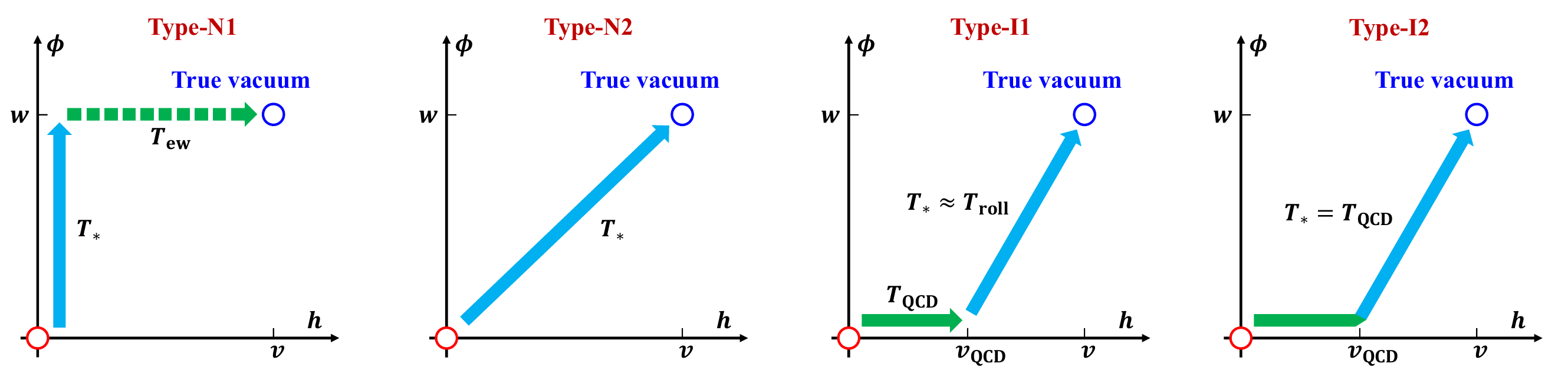}
\caption{The field space trajectories of the possible cosmological thermal history. {\bf Type-N1}: a conformal FOPT at $T_*$ followed by an EW crossover at $T_{\rm ew}$. {\bf Type-N2}: a joint conformal-EW FOPT at $T_*$. {\bf Type-I1}: a QCD-EW FOPT at $T_{\rm QCD}$ followed by a conformal FOPT at $T_*\approx T_{\rm roll}$. {\bf Type-I2}: a joint QCD-EW-conformal FOPT at $T_*= T_{\rm QCD}$.}
\label{fig:patterns}
\end{center}
\end{figure*}

The scalar potential is modified by the dense and hot plasma of the early Universe to be
\begin{multline}\label{CC_VT}
V_T(h,\phi,T)=V_1(h,\phi)+V_T^{\text{1-loop}}(h,\phi,T)\\+V_{\rm daisy}(h,\phi,T),
\end{multline}
where $T$ is the temperature, $V_T^{\text{1-loop}}$ is the one-loop thermal correction, and $V_{\rm daisy}$ is the daisy resummation. The full expression is given in Appendix~\ref{app:VT} and included in our numerical calculation. For a quick qualitative understanding of cosmic history, we use the analytical approximation
\be\label{analytical_V}
V_T(h,\phi,T)\approx V_1(h,\phi)+\frac{c_\phi T^2}{2}\phi^2+\frac{c_hT^2}{2}h^2,
\ee
where the coefficient
\be
c_h=\frac{3g^2+g'^2}{16}+\frac{y_t^2}{4}+\frac{\lambda_h}{2}\approx0.4
\ee
is caused by the SM particles, and
\be\label{c}
c_\phi=\begin{dcases}\frac{g_X^2}{4},&\text{for the gauge-induced scenario;}\\
\frac{\lambda_X}{24},&\text{for the scalar-induced scenario,}
\end{dcases}
\ee
is the BSM thermal correction coming from the heavy particle responsible for the Coleman-Weinberg potential. Although being relatively insignificant in collider studies, $Z'$ or $X$ manifests itself via the finite-temperature effects in early Universe, leading to significant consequences as detailed below.

If after inflationary reheating $T\gg w$, the $T^2$-terms dominates \Eq{analytical_V}, placing the vacuum at the origin of field space $(h,\phi)=(0,0)$, leading to symmetry restoration. As the Universe cools, the vacuum of $V_T(h,\phi,T)$ eventually transitions to the zero-temperature position $(v,w)$. A notable feature arises from the logarithmic shape of the zero-temperature potential $V_1(h,\phi)$: it is flat near the origin, with all first- and second-order derivatives vanishing. Consequently, as long as $T>0$, the $T^2$-terms induce a local minimum for $V_T(h,\phi,T)$ at the origin. This indicates that the vacuum transition from $(0,0)$ at high temperatures to $(v,w)$ at zero temperature is not a smooth roll but rather a quantum tunneling process. At a specific temperature $T_*$, $V_T(h,\phi,T_*)$ has two minima: the old vacuum $(0,0)$ and a new non-origin vacuum $(v_*,w_*)$, with the latter being the global minimum to which the Universe decays. This constitutes the cosmic FOPT, during which true vacuum bubbles nucleate, expand, and ultimately fill the entire Universe. After the FOPT, $(v_*,w_*)$ smoothly shifts to $(v,w)$ as $T\to0$.

Below the critical temperature $T_c$, the non-origin minimum becomes the global minimum (true vacuum), and the decay rate per unit volume is~\cite{Linde:1981zj}
\be
\Gamma(T)\sim T^4\left(\frac{S_3}{2\pi T}\right)^{3/2}e^{-S_3/T},
\ee
where $S_3/T$ is the action of $O(3)$-symmetric bounce solution. The false vacuum fraction of the Universe is $p(T)=e^{-I(T)}$, with~\cite{Guth:1979bh,Guth:1981uk}
\be\label{IT}
I(T)=\frac{4\pi}{3}\int_{T}^{T_c}\frac{\Gamma(T')\d T'}{T'^4H(T')}\left[\int_{T}^{T'}\frac{v_w\d T''}{H(T'')}\right]^3,
\ee
where $v_w$ is the bubble wall expansion velocity, and $H(T)$ is the Hubble constant. As the FOPT progresses, $p(T)\to0$, and the true vacuum bubbles fulfill the space. The percolation temperature $T_*$ is defined at bubbles forming an infinite connected cluster, occurring at $p(T^*) = 0.71$~\cite{rintoul1997precise}.

FOPTs in radiative symmetry breaking (i.e., classically conformal) theories have garnered significant attention~\cite{Huang:2022vkf,Chun:2023ezg,Hambye:2013dgv,Kang:2020jeg,Hambye:2018qjv,Baldes:2018emh,YaserAyazi:2019caf,Mohamadnejad:2019vzg,Khoze:2022nyt,Frandsen:2022klh,Wong:2023qon,Gouttenoire:2023pxh,Salvio:2023blb,Salvio:2023ynn,Conaci:2024tlc,Banerjee:2024fam,Witten:1980ez,Konstandin:2011dr,Jinno:2016knw,Iso:2017uuu,Ghorbani:2017lyk,Marzo:2018nov,Bian:2019szo,Ellis:2019oqb,Ellis:2020nnr,Jung:2021vap,Kawana:2022fum,Zhao:2022cnn,Sagunski:2023ynd,Ahriche:2023jdq,Ghorbani:2024twk}. This topic is particularly important because the FOPTs are in general ultra-supercooled with $T_* \ll w$, significantly impacting cosmological history. For instance, in the gauge-induced scenario, $S_3/T\propto g_X^{-3}$~\cite{Iso:2017uuu}, thus $\Gamma(T)\propto e^{-S_3/T}$ is strongly suppressed for small $g_X$. Consequently, FOPTs may occur at very late times, resulting in supercooling. Depending on the FOPT details, the evolution path of the Universe can be categorized into two main types, each with two variations, leading to four distinct possibilities.

Let $T_{\rm QCD}\approx 85$ MeV be a characteristic QCD temperature (to be explained later). When the conformal symmetry breaking FOPT occurs at $T_* > T_{\rm QCD}$, this is termed the {\it normal pattern history}. After the transition, $\phi\approx w$, and \Eq{analytical_V} reduces to
\be
V_T(h,w,T)\approx \frac12\left(c_hT^2-\frac{m_h^2}{2}\right)h^2+\frac{\lambda_h}{4}h^4.
\ee
The sign of the coefficient of $h^2$ in this potential depends on the hierarchy between $T_*$ and $T_{\rm ew}=m_h/\sqrt{2c_h}\approx140$ GeV, classifying two sub-types of evolution possibilities.
\begin{enumerate}
\item Type-N1, $T_*>T_{\rm ew}$. The EW symmetry remains preserved after the conformal FOPT. An EW crossover occurs at $T_{\rm ew}$ where $h$ shifts smoothly to $v$.
\item Type-N2, $T_*<T_{\rm ew}$. The EWSB simultaneously occurs with the conformal FOPT, resulting in a joint conformal-EW FOPT at $T_*$.
\end{enumerate}

If the decay rate is sufficiently low for the Universe to remain at $(0,0)$ until $T_{\rm QCD}$, then the QCD phase transition occurs first, a scenario we call {\it inverted pattern history}. In this case, the QCD phase transition takes place with six-flavor massless quarks, resulting in a FOPT~\cite{Braun:2006jd,Pisarski:1983ms,Guan:2024ccw}, as opposed to a crossover in the SM thermal history. The QCD also triggers an EW FOPT from $h=0$ to $v_{\rm QCD}\approx100$ MeV via the top quark condensate and Yukawa interaction $-y_th\ave{\bar tt}/\sqrt{2}$~\cite{Witten:1980ez,Iso:2017uuu}. After this joint QCD-EW FOPT, $h\approx v_{\rm QCD}$, and \Eq{analytical_V} simplifies to
\begin{multline}
V_T(v_{\rm QCD},\phi,T)\approx\frac12\left(c_\phi T^2-\frac{m_h^2v_{\rm QCD}^2}{2w^2}\right)\phi^2\\
+\frac{B}{4}\phi^4\left(\log\frac{\phi}{w}-\frac14\right).
\end{multline}
The sign of the coefficient $\phi^2$ depends on the hierarchy between $T_{\rm QCD}$ and $T_{\rm roll}=m_hv_{\rm QCD}/(\sqrt{2c_\phi}w)$, classifying two sub-types of evolution possibilities.
\begin{enumerate}
\item Type-I1, $T_{\rm QCD}>T_{\rm roll}$. After the QCD-EW FOPT, a $\phi$-direction FOPT occurs at $T_*\approx T_{\rm roll}$, which also induces the transition of $h$ from $v_{\rm QCD}$ to $\sim v$.
\item Type-I2, $T_{\rm QCD}<T_{\rm roll}$. The $\phi$-direction also gains a VEV at QCD-EW FOPT, thus this is in fact a joint QCD-EW-conformal FOPT at $T_*=T_{\rm QCD}$. 
\end{enumerate}

The field evolution trajectories of the four thermal history patterns are sketched in Fig.~\ref{fig:patterns}. The existence of the inverted pattern was proposed and studied in Refs.~\cite{Witten:1980ez, Iso:2017uuu}, while Type-I2 has been discussed in detail using low-energy QCD effective models~\cite{Sagunski:2023ynd}. Notably, each pattern includes at least one FOPT, highlighting a distinctive characteristic of radiative EWSB models. This contrasts with other BSM frameworks, where FOPTs typically occur only within specific parameter spaces and necessitate extensive parameter scanning.

It is important to note that reheating after the FOPT is not included in this discussion for simplicity. Supercooled FOPTs release a significant amount of vacuum energy into the plasma, reheating the Universe to a temperature $T_{\rm rh} \geqslant T_*$. The cosmic history is further complicated if $T_{\rm rh} \gg T_*$. For example, if $T_* < T_{\rm ew} < T_{\rm rh}$, the evolution is a Type-N2 trajectory followed by the EW symmetry restoration at $T_{\rm rh}$, and then an EW crossover at $T_{\rm ew}$.

\subsection{FOPT dynamics and GW detection}

In this research, we focus on the case of $w\gg v$ and hence $|\lambda_{h\phi}|\ll1$. Consequently, the FOPT dynamics of the $\phi$-direction can be treated separately from the SM sector, and we adopt the $\phi$-dependent thermal potential as
\be
V_T(\phi,T)\approx V_T(0,\phi,T)+V_{\rm QCD}(\phi,T),
\ee
where
\be\label{VQCD}
\delta V_{\rm QCD}(\phi,T)=\begin{dcases}~0,&T>T_{\rm QCD};\\
~-\frac{m_h^2v_{\rm QCD}^2}{4w^2}\phi^2,&T<T_{\rm QCD},\end{dcases}
\ee
is added to mimic the effect of the QCD-EW FOPT. This approach allows us to calculate the four thermal history patterns, except for Type-I2, which requires a detailed treatment of the QCD transition~\cite{Sagunski:2023ynd}. Fortunately, the parameter space of interest primarily involves Types-N1, N2, and I1, making this method sufficient for our analysis.

To calculate $\Gamma(T)$, one needs to evaluate the $O(3)$-symmetric bounce $\phi(r)$ solution by solving
\be\label{original_EoM}
\frac{\d^2\phi}{\d r^2}+\frac{2}{r}\frac{\d\phi}{\d r}=\frac{\d V_T}{\d\phi},\quad\frac{\d \phi}{\d r}\Big|_{r=0}=0,\quad \lim_{r\to\infty}\phi = 0,
\ee
where the Euclidean action is\footnote{We have confirmed that the $O(3)$-symmetric action always dominate the vacuum decay rate compared to the $O(4)$-symmetric action in the parameter space under consideration.}
\be
\frac{S_3}{T}=\frac1T\int_0^\infty4\pi r^2\d r\left[\frac12\left(\frac{\d\phi}{\d r}\right)^2+V_T\left(\phi(r),T\right)\right].
\ee
After determining $\Gamma(T)$, we derive $p(T)$ and resolve $T_*$ assuming $v_w\approx1$. If $T_*>T_{\rm QCD}$ and the FOPT completion condition~\cite{Ellis:2018mja,Turner:1992tz}
\be
3+T_*\frac{\d I}{\d T}\Big|_{T_*}<0
\ee
is satisfied, ensuring that the physical volume of the false vacuum is still decreasing at percolation, this corresponds to the normal pattern. Conversely, if $T_* < T_{\rm QCD}$, then the QCD-EW FOPT occurs, indicating the inverted pattern. We have developed and optimized homemade codes to solve \Eq{original_EoM} for $T \ll w$.

\begin{figure}
\begin{center}
\includegraphics[scale=0.375]{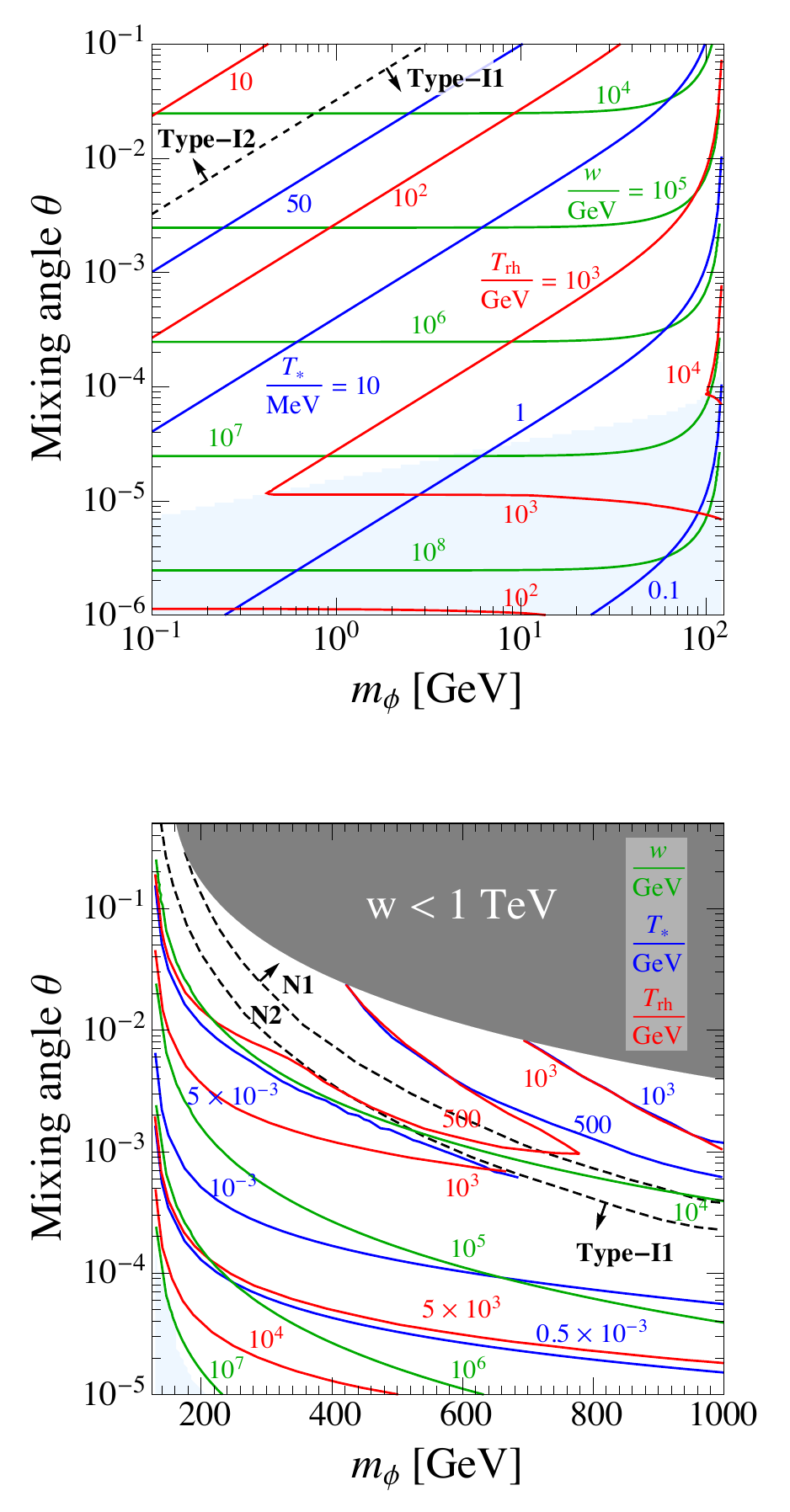}
\caption{Contours of $w$ (green) and FOPT characteristic temperatures $T_*$ (blue) and $T_{\rm rh}$ (red) for the gauge-induced scenario. The black dashed lines are the boundaries of different thermal history patterns, and the light blue shaded regions correspond to slow reheating. The top (bottom) panel is for the $m_\phi<m_h$ ($m_\phi>m_h$) case.}
\label{fig:fopt_scan}
\end{center}
\end{figure}

The reheating temperature following the FOPT is~\cite{Hambye:2018qjv}
\be\label{Trh}
T_{\rm rh}=T_\Lambda\times{\rm min}\left\{1,~\frac{\Gamma_s}{H(T_*)}\right\}^{1/2},
\ee
where $\Gamma_s=\Gamma_h\sin^2\theta+\Gamma_\phi\cos^2\theta$ with $\Gamma_h$ and $\Gamma_\phi$ being the decay width of the $h$ and $\phi$ bosons, respectively, and $T_\Lambda$ is the temperature of vacuum-radiation equality defined by
\be
\frac{\pi^2}{30}g_*(T_\Lambda)T_\Lambda^4=\Delta V_T(T_\Lambda),
\ee
with $g_*$ the effective degrees of freedom, and $\Delta V_T(T)$ the vacuum energy difference between the false and true vacua at $T$. If $\Gamma_s\gtrsim H(T_*)$, the reheating is instant, and $T_{\rm rh}=T_\Lambda$.

Fig.~\ref{fig:fopt_scan} shows the parameter space of the gauge-induced scenario, scanning over $(m_\phi, \theta)$ and plotting the contours of $w$ (green), $T_*$ (blue), and $T_{\rm rh}$ (red). Similar results for the scalar-induced scenario are obtained. Due to the non-degeneracy between $\phi$ and $h$, we display the regions where $m_\phi < m_h$ in the top panel and $m_\phi > m_h$ in the bottom panel. The boundaries of different thermal history patterns are delineated with black dashed lines. In the case of light $\phi$, we only observe inverted patterns. Here, the conformal FOPT occurs at a low temperature, approximately $T_* \sim\mO(0.1-100)~{\rm MeV}$, followed by reheating to $T_{\rm rh} \sim \mO(10-10^4)~{\rm GeV} \gg T_*$. A region of slow reheating (i.e. $\Gamma_s<H(T_*)$) is identified for $\theta \lesssim 10^{-5}$ by a light blue shaded area.

For heavy $\phi$, both normal and inverted patterns are possible, and only a narrow region in the lower-left corner exhibits slow reheating. When the FOPT reheating is prompt, $T_{\rm rh} \approx \max\{T_*,T_{\Lambda}\}$. Therefore, when supercooling is not prominent, FOPT completes during the radiation era, yielding $T_* \approx T_{\rm rh}$, which leads to the overlap of the $10^3$ GeV red and blue contours. As $m_\phi$ is fixed and $\theta$ decreases, supercooling is enhanced, resulting in a decrease in $T_*$ and an increase in $T_\Lambda$. Consequently, $T_{\rm rh}$ initially decreases before increasing, producing a cusp-shaped red contour around 500 GeV. In the inverted pattern region, $T_{\rm rh} \gg T_*$, significant reheating is obtained.

If $T_*<T_{\Lambda}$, the Universe enters a vacuum domination era before the FOPT, known as thermal inflation~\cite{Lyth:1995ka}. On the other hand, if reheating after the FOPT is slow such that $T_{\rm rh} < T_{\Lambda}$, the Universe undergoes a matter domination era after the FOPT~\cite{Dutra:2021phm}. These varied thermal history scenarios encompassing the four previously classified patterns provide a special and interesting spacetime background for addressing the longstanding puzzles in particle physics and cosmology, including generating the baryon asymmetry~\cite{Huang:2022vkf,Chun:2023ezg} and forming dark matter~\cite{Hambye:2018qjv,Baldes:2018emh,Wong:2023qon} or even sourcing primordial black holes~\cite{Gouttenoire:2023pxh,Salvio:2023blb,Salvio:2023ynn,Conaci:2024tlc,Banerjee:2024fam}.

In this research, we focus solely on the stochastic GWs generated by the FOPT. There are three main sources of the GWs: bubble collisions, sound waves, and turbulence, with their relative strengths depending on the energy budget of the transition~\cite{Espinosa:2010hh,Ellis:2019oqb,Ellis:2020nnr,Giese:2020rtr,Wang:2022lyd,Roshan:2024qnv}. If bubble walls are still in accelerating expansion at $T_*$, most of the FOPT energy is stored in the walls, leading to dominance by bubble collisions. However, if the bubble walls reach terminal velocity before percolation, the energy is primarily released to bulk motion, making sound waves the primary source. The energy budget can be evaluated by analyzing the motion of the walls, involving competition between vacuum pressure $\Delta V_T$ and the frictional force $\mP$ from particle transitions across the wall. Different results are obtained~\cite{Ellis:2019oqb,Ellis:2020nnr} for varying scaling of the resummed $(1\to n)$-splitting-induced friction, either $\propto\gamma_w$~\cite{Bodeker:2017cim,Gouttenoire:2021kjv} or $\propto\gamma_w^2$~\cite{Hoche:2020ysm}, where $\gamma_w=(1-v_w^2)^{-1/2}$ . We apply both methods and find that the projected detectable reach is not sensitive to the chosen calculation method.

The GW spectrum is defined as the energy density fraction $\Omega_{\rm gw}(f)$, which can be expressed as numerical formulae in terms of $T_*$, $v_w$, and two more effective FOPT parameters: the ratio of latent heat to the radiation energy density
\be
\alpha=\frac{1}{\pi^2g_*(T_*)T_*^4/30}\left(\Delta V_T-T\frac{\d\Delta V_T}{\d T}\right)\Big|_{T_*},
\ee
characterizing the strength of the transition, with $\alpha>1$ implying thermal inflation; the ratio of Hubble time to the FOPT duration
\be
\frac{\beta}{H_*}=T_*\frac{\d}{\d T}\left(\frac{S_3}{T}\right)\Big|_{T_*}.
\ee
In certain parameter space regions, the FOPT is very slow and we switch to use another definition~\cite{Kanemura:2024pae}
\be
\frac{\beta}{H_*}\to\frac{(8\pi)^{1/3}v_w}{H_*\bar R},
\ee
where $\bar R$ is the mean separation of bubble~\cite{Ellis:2018mja}, which can be taken as $n_b^{-1/3}$, with $n_b$ being the bubble density~\cite{Wang:2020jrd}.

We derive $\alpha$ and $\beta/H_*$ at $T_*$ to get the GW spectra at the FOPT~\cite{Huber:2008hg,Hindmarsh:2015qta,Caprini:2009yp}, and assume that the shapes remain unchanged during the instant reheating from $T_*$ to $T_{\rm rh}$.\footnote{The numerical formula for the sound wave contribution is derived from a moderate FOPT~\cite{Hindmarsh:2015qta}. In the case of an ultra-strong FOPT with $\alpha \gg 1$, the sound wave spectrum resembles that of bubble collisions~\cite{Lewicki:2022pdb,Ellis:2023oxs,Caprini:2024hue}. However, as we will demonstrate, the parameter $\beta/H_*$ primarily determines the detectability of GW signals from ultra-supercooled FOPTs, and hence our results are insensitive to these spectral differences.} We then redshift the spectra from $T_{\rm rh}$ to today $T_0\approx2.73$ K~\cite{Breitbach:2018ddu}. Note that if $T_{\rm rh}\approx T_*$ then the treatment is equivalent to the numerical formulae in Refs.~\cite{Caprini:2015zlo,Caprini:2019egz}. However, in our scenario, usually $T_{\rm rh}\gg T_*$, thus the difference is significant. If the reheating is slow, $T_{\rm rh}\ll T_\Lambda$, the GW shape is further affected~\cite{Ellis:2020nnr},  but this is not a concern in the GW-detectable parameter space. The GW spectra today lie within the sensitivity region of the future space-based interferometer GW detectors such as LISA~\cite{LISA:2017pwj}, TianQin~\cite{TianQin:2015yph}, Taiji~\cite{Hu:2017mde}, and BBO~\cite{Crowder:2005nr}.

\begin{figure}
\begin{center}
\includegraphics[scale=0.35]{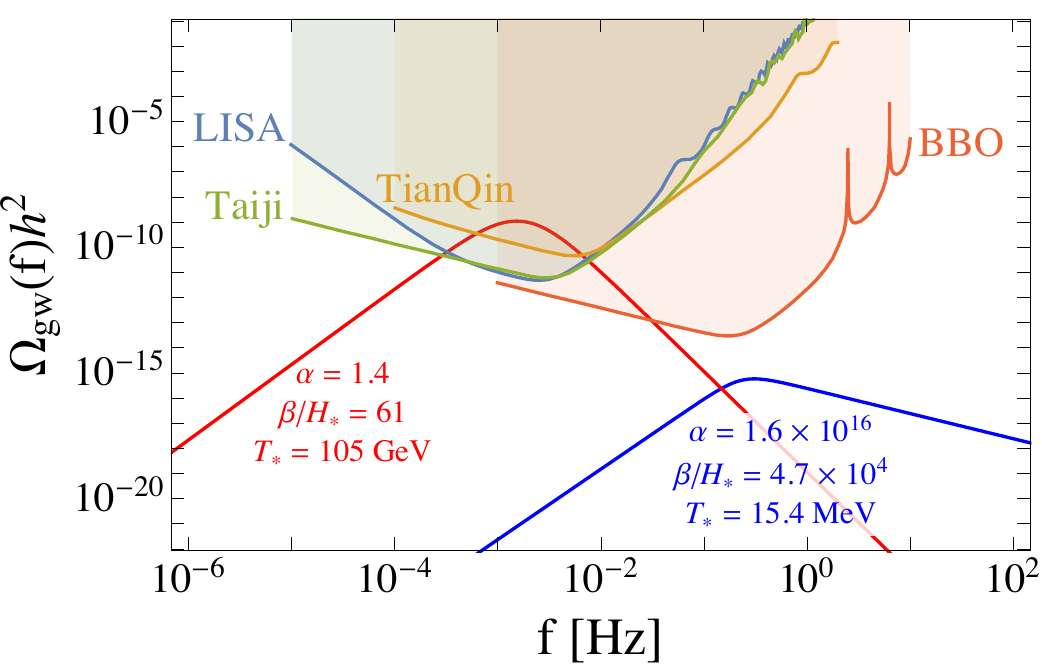}
\caption{Illustrations of the GW spectra for $m_\phi=200$ GeV, $\theta=0.1$ ($w=1.6$ TeV and $g_X=0.79$, red curve) and $m_\phi=10$ GeV, $\theta=0.01$ ($w=25$ TeV and $g_X=0.046$, blue curve) in the gauge-induced scenario, with the FOPT parameters given in the figure.}
\label{fig:GW_spectrum}
\end{center}
\end{figure}

\begin{figure*}[t]
\begin{center}
\includegraphics[scale=0.375]{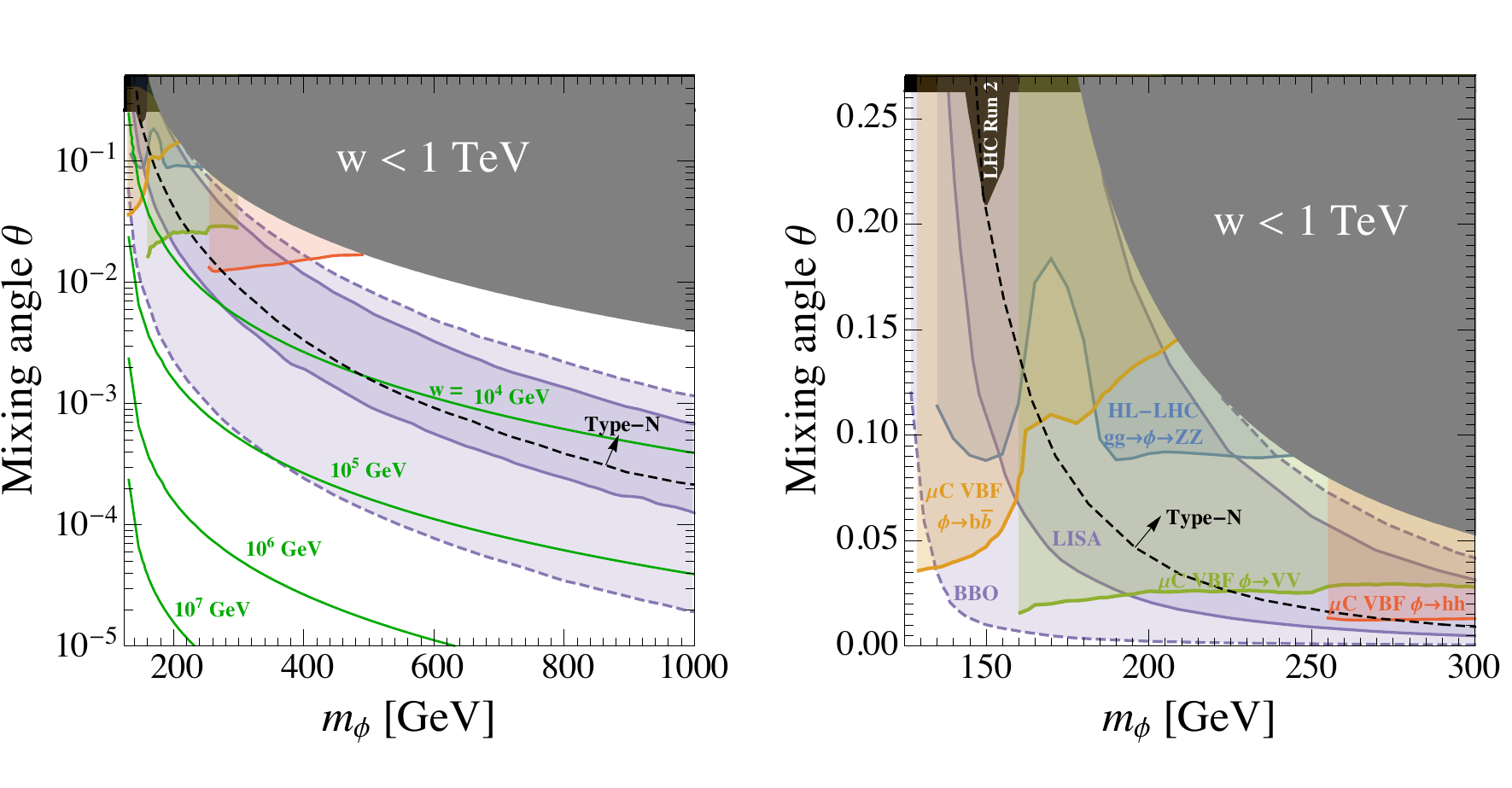}
\caption{Current bounds and projected limits on the parameter space for the $m_\phi>m_h$ case, with the right panel being the zoomed-in figure of the left panel in the mass range $m_\phi\in[m_h,~300~{\rm GeV}]$. The black shaded region is already excluded, while the colored shaded regions represent the reach of future experiments. The black dashed lines are the boundaries between normal and inverted thermal history patterns, while the green solid lines plot the contour of $w$.}
\label{fig:sup}
\end{center}
\end{figure*}

Before presenting the projected reach in the next section, we briefly comment on the GW spectra. Most parameter space reveals strong FOPTs with $\alpha\gg1$. However, this doesn't necessarily imply strong GWs. As noted in Ref.~\cite{Kanemura:2024pae}, strong GWs require a transition that is both strong and slow, characterized by large $\alpha$ and small $\beta/H_*$. In the radiative EWSB model, many regions allow for ultra-supercooled FOPTs with prolonged thermal inflation, greatly impacting cosmic history but resulting in rapid transitions that produce no detectable GWs. To illustrate, Fig.~\ref{fig:GW_spectrum} shows the GW spectra for two benchmarks in the gauge-induced radiative symmetry breaking scenario. The red curve corresponds to a strong, slow transition with detectable signals for instruments like LISA; while the blue curve represents a super-strong but prompt transition, yielding signals too weak for detection, even though $\alpha\sim10^{16}$ is extremely large.

\section{Results}\label{sec:results}

Combining the discussions in Section~\ref{sec:collider} and Section~\ref{sec:GW}, this section presents the main results of our research. Fig.~\ref{fig:sup} displays current bounds and projected limits for the heavy scalar case where $m_\phi > m_h$. The left panel shows a scan of $m_\phi$ over the range $[m_h,~1~{\rm TeV}]$ (linear scale) and $\theta$ over $[10^{-6},~{0.5}]$ (log scale). The right panel provides a detailed view of the region $m_\phi\in[m_h,~300~{\rm GeV}]$ and $\theta\in[0,~0.27]$ (double-linear scale). Due to the validity of our sequential symmetry breaking treatment for $w \gg v$, we exclude the parameter space where $w<1$ TeV with the gray shaded region.

The current bounds are derived from LHC Run 2 results on Higgs signal strength~\cite{ATLAS:2024fkg} (combining $36.1~{\rm fb}^{-1}$ to $139~{\rm fb}^{-1}$) and BSM searches for $\phi \to ZZ$~\cite{CMS:2018amk} (with $35.9~{\rm fb}^{-1}$). The colored shaded regions indicate various future projections. The CMS $\phi \to ZZ$ result is rescaled to $3000~{\rm fb}^{-1}$ for the HL-LHC reach, which can achieve sensitivity of $\theta \sim 0.1$ for $m_h \lesssim m_\phi \lesssim 2m_h$ when the $\phi\to ZZ^{(*)}$ branching ratio is significant. The reduced sensitivity around $m_\phi \sim 170$ GeV is due to the suppression of the branching ratio~\cite{Djouadi:2005gi}. Additionally, projections for VBF $\phi\to b\bar{b}$, $VV$, and $hh$ channels at future 10 TeV muon colliders indicate sensitivities reaching $\theta \sim 10^{-2}$. The dominance of different channels across various mass ranges reflects the branching ratio characteristics described in \Eq{brs}.

We present projections for future GW detectors LISA~\cite{LISA:2017pwj} and its proposed successor BBO~\cite{Crowder:2005nr}. TianQin~\cite{TianQin:2015yph} and Taiji~\cite{Hu:2017mde} are expected to yield similar sensitivity with LISA. The detection limit is defined by requiring the signal-to-noise ratio
\be
{\rm SNR}=\sqrt{\mT\int\d f\left(\frac{\Omega_{\rm gw}(f)}{\Omega_{\rm detector}(f)}\right)^2}=50,
\ee
using the sensitivity curves $\Omega_{\rm detector}(f)$ of LISA and BBO from Ref.~\cite{Schmitz:2020syl}, and the operational time is approximately $\mT\approx 9.46\times10^7 ~{\rm s}=0.75\times4$ years~\cite{Caprini:2019egz}. The projected region for GW detection is significantly larger than that for colliders in the log-$\theta$ coordinate perspective, with LISA probing $\theta$ down to $10^{-4}$ and BBO down to $10^{-5}$, and $w$ up to $10^4$ GeV and $10^5$ GeV, respectively.

While the collider reach is independent of the origin of radiative symmetry breaking, the dynamics of FOPT does depend on the origin of the parameter $B$. Consequently, gauge- and scalar-induced scenarios yield different GW spectra for a given parameter point $(m_\phi, \theta)$, and here we show the results for the former scenario. However, we find that the projected reach of $\theta$ for a given $m_\phi$ varies by less than a factor of 2, which is negligible compared to the uncertainties in FOPT GW calculations~\cite{Athron:2023rfq,Guo:2021qcq}. Therefore, the probed region in Fig.~\ref{fig:sup} is insensitive to the origin of $B$. Henceforth, we will use the gauge-induced scenario as our primary example.

The left panel of Fig.~\ref{fig:sup} illustrates that GW and collider searches are complementary across most of the parameter space. The right panel zooms in on the region where these searches overlap, allowing for crosscheck. A future GW excess detected by LISA, TianQin, Taiji, or BBO could be further validated by signals from the HL-LHC or muon collider, supporting its origin as a FOPT via the radiative symmetry breaking mechanism. As shown in the figure, the $\phi \to hh$ channel covers only a small portion of the GW detection region, whereas the $\phi \to b\bar{b}$ and $\phi \to VV$ channels provide significant contributions for verifying GW detection. This is different from the xSM, where the majority of the GW detectable parameter space can be probed by the $\phi \to hh$ channel at the 10 TeV muon collider~\cite{Liu:2021jyc} (see also Refs.~\cite{Alves:2018oct,Alves:2019igs,Alves:2020bpi} for di-Higgs probes of xSM FOPTs).

\begin{figure}
\begin{center}
\includegraphics[scale=0.375]{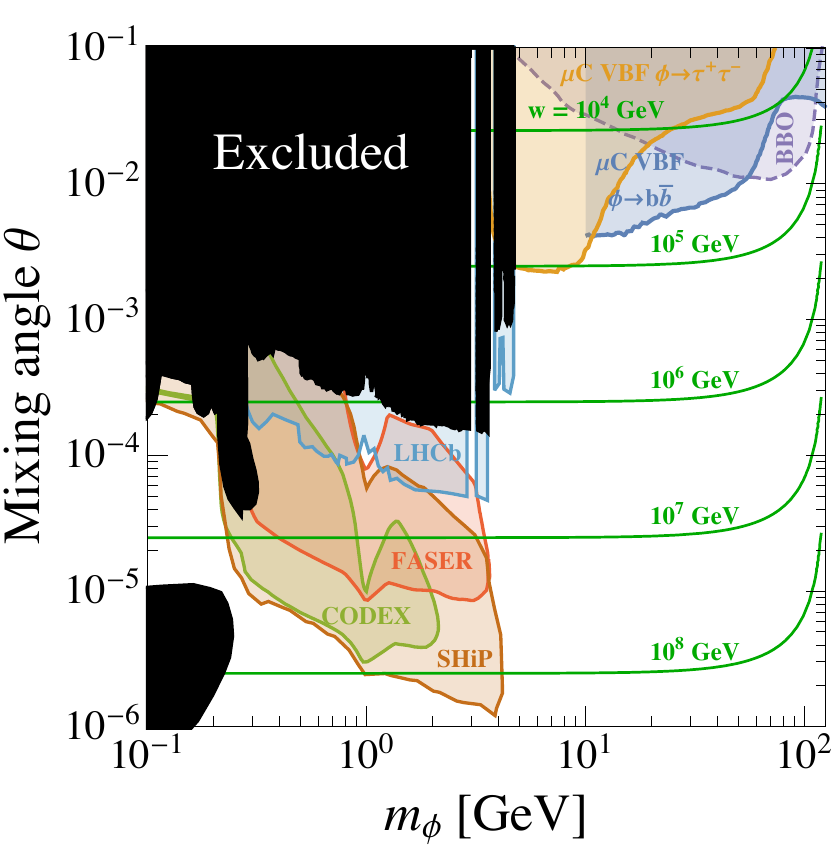}
\caption{Current bounds and projected limits on the parameter space for the $m_\phi<m_h$ case. The black shaded region is already excluded, while the colored shaded regions represent the reach of future experiments. The green solid lines plot the contour of $w$.}
\label{fig:sub}
\end{center}
\end{figure}

The combined results for the $m_\phi < m_h$ case are shown in Fig.~\ref{fig:sub} for $m_\phi\in[10^{-1}~{\rm GeV},~m_h]$ and $\theta\in[10^{-6},~10^{-1}]$ (double-log scale), where the black region is excluded, and the colored shaded areas indicate future projections. The green solid lines represent contours of $w$. Current bounds arise from various collider and beam-dump experiments that search for a light scalar boson mixing with the Higgs boson, including LHCb~\cite{LHCb:2016awg,LHCb:2015nkv}, NA62~\cite{NA62:2020pwi,NA62:2020xlg}, CHARM~\cite{Winkler:2018qyg}, E949~\cite{BNL-E949:2009dza}, LSND~\cite{Foroughi-Abari:2020gju}, and MicroBooNE~\cite{MicroBooNE:2020vlq}. The constraint from SN1987 is illustrated as a separate shaded region in the bottom-left corner~\cite{Winkler:2018qyg}. We also show the projected reach for $m_\phi\lesssim2m_\tau$ from the LLP searches in LHCb, FASER~\cite{FASER:2018eoc}, CODEX~\cite{Gligorov:2017nwh}, and SHiP~\cite{SHiP:2020vbd} based on results from Refs.~\cite{Feng:2017vli,Kling:2021fwx}. These future experiments can probe $\theta$ as low as $10^{-6}$ with $w$ reaching up to $10^8$ GeV.

The prompt decay of $\phi\to\tau^+\tau^-$ and $b\bar b$ can probe $m_\phi \gtrsim 2m_\tau$ with $\theta$ reaching a few $10^{-3}$. Notably, this region can be crosschecked by signals detected by BBO. As shown in Fig.~\ref{fig:fopt_scan} of Section~\ref{sec:GW}, the $m_\phi < m_h$ case corresponds to an inverted thermal history where the conformal phase transition is delayed until after the QCD-EW FOPT. The Type-I2 region is in the top-left part of the figure and excluded by existing data, leaving the viable type-I1 region, where a $\phi$-direction FOPT occurs at $T_* \approx T_{\text{roll}} < T_{\rm QCD}$. While such a FOPT can be extremely strong, with $\alpha$ reaching up to $10^{30}$, the duration of the transition is very short, yielding $\beta / H_*$ up to $10^9$. As a result, the GWs produced are weak, and only BBO can explore a small fraction of the parameter space in the top-right corner of the figure. This counterintuitive result arises because, following the QCD-EW FOPT, a negative mass squared term \Eq{VQCD} is induced along the $\phi$-direction, causing the local minimum of $\phi=0$ to disappear at $T_{\rm roll}$ and resulting in a rapid transition.

We note that Higgs exotic decay $h\to\phi\phi$ is not sensitive to the radiative symmetry breaking mechanism, as the branching ratio $\lesssim10^{-10}$ in the $m_\phi<m_h/2$ region of Fig.~\ref{fig:sub}. This suppression comes from the $m_\phi^2/w^2$ factor in the coefficient of the $h\phi^2$ term in \Eq{L3}. This behavior is a characteristic of the radiative symmetry breaking scenario and contrasts significantly with non-conformal extensions of the SM, such as the $U(1)'$-extension or xSM, where Higgs exotic decays can effectively probe FOPTs~\cite{Li:2023bxy,Carena:2022yvx,Kozaczuk:2019pet,Carena:2019une,Liu:2022nvk,Kanemura:2023jiw}.

\section{Conclusion}\label{sec:conclusion}

We have presented a detailed phenomenological analysis of radiative EWSB, focusing on its key feature: the logarithmic potential. Excitations of the field quanta around the vacuum yield a new scalar particle, which can be investigated at particle colliders or beam-dump experiments. Moreover, the flat potential near the origin can induce one or more FOPTs during cosmic evolution, generating stochastic GW detectable by space-based interferometers. Following a detailed analysis of the vacuum structure of the joint scalar potential, the experimental signals are studied. In collider studies, we analyze LLP signals and the prompt decay of the new scalar into SM particle pairs. On the cosmological side, we investigate FOPT dynamics, classifying four distinct thermal history patterns (with further variations due to reheating effects), and subsequently calculate the resulting GWs.

We project the results on the $(m_\phi,\theta)$ plane to provide an overview of the phenomenology. The combined results from particle and GW experiments effectively probe the parameter space, revealing both complementary and overlapping regions. For $m_\phi < m_h$, the FOPTs are always ultra-supercooled to be in the inverted pattern, however the GWs produced are not very strong due to the short phase transition duration. As a result, particle experiments are essential for effectively probing the mechanism in this case. Future LLP searches offer the most sensitive exploration of the mechanism, reaching scales up to $w \sim 10^8$ GeV and mixing angle to $\theta\sim10^{-6}$. In the case of $m_\phi > m_h$, GWs from FOPTs are strong enough to probe $w$ up to $10^5$ GeV and $\theta\sim10^{-5}$ at the BBO. In both scenarios, there is overlap region between collider and GW experiments, enabling crosschecks that can help identify the radiative symmetry breaking mechanism. Remarkably, the $\tau^+\tau^-$, $b\bar b$, $VV$, and $hh$ channels all significantly contribute to the cross-verification of GW detections, in contrast to the $hh$-dominance observed in the non-conformal xSM case.

Utilizing the diverse thermal history patterns identified, novel solutions can be proposed to BSM puzzles in particle physics and cosmology. For instance, dark matter or baryon asymmetry may be generated via particle interactions with ultra-relativistic bubble walls~\cite{Huang:2022vkf,Chun:2023ezg} or after thermal inflation~\cite{Hambye:2018qjv,Baldes:2018emh,Wong:2023qon}. Slow transition may form primordial black holes through false vacuum islands~\cite{Gouttenoire:2023pxh,Salvio:2023blb,Salvio:2023ynn,Conaci:2024tlc,Banerjee:2024fam}. See Refs.~\cite{Baldes:2021vyz,Azatov:2021irb,Azatov:2021ifm,Baldes:2023fsp,Ai:2024ikj,Liu:2021svg,Lewicki:2024ghw,Cai:2024nln} for more relevant studies based on the general supercooled FOPTs, which naturally apply to radiative symmetry breaking models. Our work can thus serve as a foundation for these further investigations.

{\bf Note added.} When calculating the false vacuum fraction, we omitted the initial bubble radius $R_0$ in \Eq{IT}. It turns out that in an ultra-supercooled type-I FOPT, this term is not negligible; including it significantly enhances the GW signals over a large part of the parameter space, making them detectable by LISA or BBO, as recently shown in Ref.~\cite{Xie:2026vor}.

\section*{Acknowledgements}

We would like to thank Shao-Ping Li for very useful discussions. W. L. is supported by National Natural Science Foundation of China (Grant No.12205153). K.-P.X. is supported by the National Natural Science Foundation of China under Grant No. 12305108. K.-P.X. is also supported by the Fundamental Research Funds for the Central Universities.

\appendix

\section{Detailed expressions of the thermal potential}\label{app:VT}

The one-loop thermal term is given by
\be\label{VT1loop}
V_T^{\text{1-loop}}(h,\phi,T)=\sum_i\frac{n_iT^4}{2\pi^2}J_{B/F}\left(\frac{M_i^2}{T^2}\right),
\ee
where the summation runs over particles whose mass depend on the background fields $h$ and $\phi$, and the thermal integrations are defined as
\be
J_{B/F}(y)=\pm\int_0^\infty x^2{\rm d}x\ln\left(1\mp e^{-\sqrt{x^2+y}}\right),
\ee
with the subscript $B$ and $F$ denoting bosonic and fermionic contribution, respectively. The field-dependent masses and numbers of effective degrees of freedom are
\be
\begin{dcases}~M_W(h)=\frac{g}{2}h,&n_W=2\times 3=6;\\
~M_Z(h)=\frac{\sqrt{g^2+g'^2}}{2}h,&n_Z=3;\\
~M_t(h)=\frac{y_t}{\sqrt{2}}h,&n_t=N_c\times4=12,
\end{dcases}
\ee
for the SM gauge bosons and top quark where $g$ and $g'$ are the gauge couplings of the SM $SU(2)_L$ and $U(1)_Y$ groups, respectively, and $y_t$ is the top quark Yukawa. The SM scalars have
\be\begin{dcases}~M_h^2(h,\phi)=3\lambda_hh^2+\frac{\lambda_{h\phi}}{2}\phi^2,& n_h=1;\\
M_G^2(h,\phi)=\lambda_hh^2+\frac{\lambda_{h\phi}}{2}\phi^2,& n_G=3.
\end{dcases}\ee
The BSM sector has
\be\label{MZ}
M_{Z'}(\phi)=g_X\phi,\quad n_{Z'}=3,
\ee
for the gauge-induced scenario and
\be\label{MX}
M_X(\phi)=\sqrt{\frac{\lambda_X}{2}}\phi,\quad n_X=1,
\ee
for the scalar-induced scenario.

The daisy resummation term can be decomposed as to the SM and BSM components, and the latter is
\be\label{BSMdaisy}
\begin{dcases}~-\frac{T}{12\pi}g_X^3\left[\left(\phi^2+T^2\right)^{3/2}-\phi^3\right],\\
~-\frac{T}{12\pi}\left(\frac{\lambda_X}{2}\right)^{3/2}\left[\left(\phi^2+\frac{T^2}{12}\right)^{3/2}-\phi^3\right],
\end{dcases}
\ee
for the gauge- and scalar-induced radiative symmetry breaking, respectively. The expression of the SM daisy resummation is involved and not crucial for our discussion, thus is not shown here, and we refer the readers to Ref.~\cite{Carrington:1991hz} for the details.

The approximate potential \Eq{analytical_V} is obtained by expanding the thermal integrations around $y\approx0$:
\be\begin{split}
J_B(y)\approx&-\frac{\pi^4}{45}+\frac{\pi^2}{12}y-\frac{\pi}{6}y^{3/2}-\frac{y^2}{32}\log\frac{y}{a_B},\\
J_F(y)\approx&-\frac{7\pi^4}{360}+\frac{\pi^2}{24}y+\frac{y^2}{32}\log\frac{y}{a_F},
\end{split}\ee
where $a_B=16a_F$ and $a_F=\pi^2e^{1.5-2\gamma_E}$ with $\gamma_E\approx0.577$ the Euler's constant.

\bibliographystyle{apsrev}
\bibliography{references}

\begin{thebibliography}{132}
\expandafter\ifx\csname natexlab\endcsname\relax\def\natexlab#1{#1}\fi
\expandafter\ifx\csname bibnamefont\endcsname\relax
  \def\bibnamefont#1{#1}\fi
\expandafter\ifx\csname bibfnamefont\endcsname\relax
  \def\bibfnamefont#1{#1}\fi
\expandafter\ifx\csname citenamefont\endcsname\relax
  \def\citenamefont#1{#1}\fi
\expandafter\ifx\csname url\endcsname\relax
  \def\url#1{\texttt{#1}}\fi
\expandafter\ifx\csname urlprefix\endcsname\relax\def\urlprefix{URL }\fi
\providecommand{\bibinfo}[2]{#2}
\providecommand{\eprint}[2][]{\url{#2}}

\bibitem[{\citenamefont{Aad et~al.}(2012)}]{ATLAS:2012yve}
\bibinfo{author}{\bibfnamefont{G.}~\bibnamefont{Aad}} \bibnamefont{et~al.}
  (\bibinfo{collaboration}{ATLAS}), \bibinfo{journal}{Phys. Lett. B}
  \textbf{\bibinfo{volume}{716}}, \bibinfo{pages}{1} (\bibinfo{year}{2012}),
  \eprint{1207.7214}.

\bibitem[{\citenamefont{Chatrchyan et~al.}(2012)}]{CMS:2012qbp}
\bibinfo{author}{\bibfnamefont{S.}~\bibnamefont{Chatrchyan}}
  \bibnamefont{et~al.} (\bibinfo{collaboration}{CMS}), \bibinfo{journal}{Phys.
  Lett. B} \textbf{\bibinfo{volume}{716}}, \bibinfo{pages}{30}
  (\bibinfo{year}{2012}), \eprint{1207.7235}.

\bibitem[{\citenamefont{Coleman and Weinberg}(1973)}]{Coleman:1973jx}
\bibinfo{author}{\bibfnamefont{S.~R.} \bibnamefont{Coleman}} \bibnamefont{and}
  \bibinfo{author}{\bibfnamefont{E.~J.} \bibnamefont{Weinberg}},
  \bibinfo{journal}{Phys. Rev. D} \textbf{\bibinfo{volume}{7}},
  \bibinfo{pages}{1888} (\bibinfo{year}{1973}).

\bibitem[{\citenamefont{Jackiw}(1974)}]{Jackiw:1974cv}
\bibinfo{author}{\bibfnamefont{R.}~\bibnamefont{Jackiw}},
  \bibinfo{journal}{Phys. Rev. D} \textbf{\bibinfo{volume}{9}},
  \bibinfo{pages}{1686} (\bibinfo{year}{1974}).

\bibitem[{\citenamefont{Bardeen}(1995)}]{Bardeen:1995kv}
\bibinfo{author}{\bibfnamefont{W.~A.} \bibnamefont{Bardeen}}, in
  \emph{\bibinfo{booktitle}{{Ontake Summer Institute on Particle Physics}}}
  (\bibinfo{year}{1995}).

\bibitem[{\citenamefont{Meissner and Nicolai}(2008)}]{Meissner:2007xv}
\bibinfo{author}{\bibfnamefont{K.~A.} \bibnamefont{Meissner}} \bibnamefont{and}
  \bibinfo{author}{\bibfnamefont{H.}~\bibnamefont{Nicolai}},
  \bibinfo{journal}{Phys. Lett. B} \textbf{\bibinfo{volume}{660}},
  \bibinfo{pages}{260} (\bibinfo{year}{2008}), \eprint{0710.2840}.

\bibitem[{\citenamefont{de~Boer et~al.}(2024)\citenamefont{de~Boer, Lindner,
  and Trautner}}]{deBoer:2024jne}
\bibinfo{author}{\bibfnamefont{T.}~\bibnamefont{de~Boer}},
  \bibinfo{author}{\bibfnamefont{M.}~\bibnamefont{Lindner}}, \bibnamefont{and}
  \bibinfo{author}{\bibfnamefont{A.}~\bibnamefont{Trautner}}
  (\bibinfo{year}{2024}), \eprint{2407.15920}.

\bibitem[{\citenamefont{Frasca et~al.}(2024)\citenamefont{Frasca, Ghoshal, and
  Okada}}]{Frasca:2024fuq}
\bibinfo{author}{\bibfnamefont{M.}~\bibnamefont{Frasca}},
  \bibinfo{author}{\bibfnamefont{A.}~\bibnamefont{Ghoshal}}, \bibnamefont{and}
  \bibinfo{author}{\bibfnamefont{N.}~\bibnamefont{Okada}}
  (\bibinfo{year}{2024}), \eprint{2408.00093}.

\bibitem[{\citenamefont{Nakayama}(2015)}]{Nakayama:2013is}
\bibinfo{author}{\bibfnamefont{Y.}~\bibnamefont{Nakayama}},
  \bibinfo{journal}{Phys. Rept.} \textbf{\bibinfo{volume}{569}},
  \bibinfo{pages}{1} (\bibinfo{year}{2015}), \eprint{1302.0884}.

\bibitem[{\citenamefont{Hempfling}(1996)}]{Hempfling:1996ht}
\bibinfo{author}{\bibfnamefont{R.}~\bibnamefont{Hempfling}},
  \bibinfo{journal}{Phys. Lett. B} \textbf{\bibinfo{volume}{379}},
  \bibinfo{pages}{153} (\bibinfo{year}{1996}), \eprint{hep-ph/9604278}.

\bibitem[{\citenamefont{Iso et~al.}(2009{\natexlab{a}})\citenamefont{Iso,
  Okada, and Orikasa}}]{Iso:2009ss}
\bibinfo{author}{\bibfnamefont{S.}~\bibnamefont{Iso}},
  \bibinfo{author}{\bibfnamefont{N.}~\bibnamefont{Okada}}, \bibnamefont{and}
  \bibinfo{author}{\bibfnamefont{Y.}~\bibnamefont{Orikasa}},
  \bibinfo{journal}{Phys. Lett. B} \textbf{\bibinfo{volume}{676}},
  \bibinfo{pages}{81} (\bibinfo{year}{2009}{\natexlab{a}}), \eprint{0902.4050}.

\bibitem[{\citenamefont{Iso et~al.}(2009{\natexlab{b}})\citenamefont{Iso,
  Okada, and Orikasa}}]{Iso:2009nw}
\bibinfo{author}{\bibfnamefont{S.}~\bibnamefont{Iso}},
  \bibinfo{author}{\bibfnamefont{N.}~\bibnamefont{Okada}}, \bibnamefont{and}
  \bibinfo{author}{\bibfnamefont{Y.}~\bibnamefont{Orikasa}},
  \bibinfo{journal}{Phys. Rev. D} \textbf{\bibinfo{volume}{80}},
  \bibinfo{pages}{115007} (\bibinfo{year}{2009}{\natexlab{b}}),
  \eprint{0909.0128}.

\bibitem[{\citenamefont{Chun et~al.}(2013)\citenamefont{Chun, Jung, and
  Lee}}]{Chun:2013soa}
\bibinfo{author}{\bibfnamefont{E.~J.} \bibnamefont{Chun}},
  \bibinfo{author}{\bibfnamefont{S.}~\bibnamefont{Jung}}, \bibnamefont{and}
  \bibinfo{author}{\bibfnamefont{H.~M.} \bibnamefont{Lee}},
  \bibinfo{journal}{Phys. Lett. B} \textbf{\bibinfo{volume}{725}},
  \bibinfo{pages}{158} (\bibinfo{year}{2013}), \bibinfo{note}{[Erratum:
  Phys.Lett.B 730, 357--359 (2014)]}, \eprint{1304.5815}.

\bibitem[{\citenamefont{Das et~al.}(2017)\citenamefont{Das, Okada, and
  Papapietro}}]{Das:2015nwk}
\bibinfo{author}{\bibfnamefont{A.}~\bibnamefont{Das}},
  \bibinfo{author}{\bibfnamefont{N.}~\bibnamefont{Okada}}, \bibnamefont{and}
  \bibinfo{author}{\bibfnamefont{N.}~\bibnamefont{Papapietro}},
  \bibinfo{journal}{Eur. Phys. J. C} \textbf{\bibinfo{volume}{77}},
  \bibinfo{pages}{122} (\bibinfo{year}{2017}), \eprint{1509.01466}.

\bibitem[{\citenamefont{Khoze and Ro}(2013)}]{Khoze:2013oga}
\bibinfo{author}{\bibfnamefont{V.~V.} \bibnamefont{Khoze}} \bibnamefont{and}
  \bibinfo{author}{\bibfnamefont{G.}~\bibnamefont{Ro}}, \bibinfo{journal}{JHEP}
  \textbf{\bibinfo{volume}{10}}, \bibinfo{pages}{075} (\bibinfo{year}{2013}),
  \eprint{1307.3764}.

\bibitem[{\citenamefont{Davoudiasl and Lewis}(2014)}]{Davoudiasl:2014pya}
\bibinfo{author}{\bibfnamefont{H.}~\bibnamefont{Davoudiasl}} \bibnamefont{and}
  \bibinfo{author}{\bibfnamefont{I.~M.} \bibnamefont{Lewis}},
  \bibinfo{journal}{Phys. Rev. D} \textbf{\bibinfo{volume}{90}},
  \bibinfo{pages}{033003} (\bibinfo{year}{2014}), \eprint{1404.6260}.

\bibitem[{\citenamefont{Huang and Xie}(2022)}]{Huang:2022vkf}
\bibinfo{author}{\bibfnamefont{P.}~\bibnamefont{Huang}} \bibnamefont{and}
  \bibinfo{author}{\bibfnamefont{K.-P.} \bibnamefont{Xie}},
  \bibinfo{journal}{JHEP} \textbf{\bibinfo{volume}{09}}, \bibinfo{pages}{052}
  (\bibinfo{year}{2022}), \eprint{2206.04691}.

\bibitem[{\citenamefont{Chun et~al.}(2023)\citenamefont{Chun, Dutka, Jung,
  Nagels, and Vanvlasselaer}}]{Chun:2023ezg}
\bibinfo{author}{\bibfnamefont{E.~J.} \bibnamefont{Chun}},
  \bibinfo{author}{\bibfnamefont{T.~P.} \bibnamefont{Dutka}},
  \bibinfo{author}{\bibfnamefont{T.~H.} \bibnamefont{Jung}},
  \bibinfo{author}{\bibfnamefont{X.}~\bibnamefont{Nagels}}, \bibnamefont{and}
  \bibinfo{author}{\bibfnamefont{M.}~\bibnamefont{Vanvlasselaer}},
  \bibinfo{journal}{JHEP} \textbf{\bibinfo{volume}{09}}, \bibinfo{pages}{164}
  (\bibinfo{year}{2023}), \eprint{2305.10759}.

\bibitem[{\citenamefont{Okada and Orikasa}(2012)}]{Okada:2012sg}
\bibinfo{author}{\bibfnamefont{N.}~\bibnamefont{Okada}} \bibnamefont{and}
  \bibinfo{author}{\bibfnamefont{Y.}~\bibnamefont{Orikasa}},
  \bibinfo{journal}{Phys. Rev. D} \textbf{\bibinfo{volume}{85}},
  \bibinfo{pages}{115006} (\bibinfo{year}{2012}), \eprint{1202.1405}.

\bibitem[{\citenamefont{Hambye and Strumia}(2013)}]{Hambye:2013dgv}
\bibinfo{author}{\bibfnamefont{T.}~\bibnamefont{Hambye}} \bibnamefont{and}
  \bibinfo{author}{\bibfnamefont{A.}~\bibnamefont{Strumia}},
  \bibinfo{journal}{Phys. Rev. D} \textbf{\bibinfo{volume}{88}},
  \bibinfo{pages}{055022} (\bibinfo{year}{2013}), \eprint{1306.2329}.

\bibitem[{\citenamefont{Kang and Zhu}(2020)}]{Kang:2020jeg}
\bibinfo{author}{\bibfnamefont{Z.}~\bibnamefont{Kang}} \bibnamefont{and}
  \bibinfo{author}{\bibfnamefont{J.}~\bibnamefont{Zhu}},
  \bibinfo{journal}{Phys. Rev. D} \textbf{\bibinfo{volume}{102}},
  \bibinfo{pages}{053011} (\bibinfo{year}{2020}), \eprint{2003.02465}.

\bibitem[{\citenamefont{Yaser~Ayazi and
  Mohamadnejad}(2019)}]{YaserAyazi:2019caf}
\bibinfo{author}{\bibfnamefont{S.}~\bibnamefont{Yaser~Ayazi}} \bibnamefont{and}
  \bibinfo{author}{\bibfnamefont{A.}~\bibnamefont{Mohamadnejad}},
  \bibinfo{journal}{JHEP} \textbf{\bibinfo{volume}{03}}, \bibinfo{pages}{181}
  (\bibinfo{year}{2019}), \eprint{1901.04168}.

\bibitem[{\citenamefont{Mohamadnejad}(2020)}]{Mohamadnejad:2019vzg}
\bibinfo{author}{\bibfnamefont{A.}~\bibnamefont{Mohamadnejad}},
  \bibinfo{journal}{Eur. Phys. J. C} \textbf{\bibinfo{volume}{80}},
  \bibinfo{pages}{197} (\bibinfo{year}{2020}), \eprint{1907.08899}.

\bibitem[{\citenamefont{Khoze and Milne}(2023)}]{Khoze:2022nyt}
\bibinfo{author}{\bibfnamefont{V.~V.} \bibnamefont{Khoze}} \bibnamefont{and}
  \bibinfo{author}{\bibfnamefont{D.~L.} \bibnamefont{Milne}},
  \bibinfo{journal}{Phys. Rev. D} \textbf{\bibinfo{volume}{107}},
  \bibinfo{pages}{095012} (\bibinfo{year}{2023}), \eprint{2212.04784}.

\bibitem[{\citenamefont{Frandsen et~al.}(2023)\citenamefont{Frandsen,
  Heikinheimo, Thing, Tuominen, and Rosenlyst}}]{Frandsen:2022klh}
\bibinfo{author}{\bibfnamefont{M.~T.} \bibnamefont{Frandsen}},
  \bibinfo{author}{\bibfnamefont{M.}~\bibnamefont{Heikinheimo}},
  \bibinfo{author}{\bibfnamefont{M.~E.} \bibnamefont{Thing}},
  \bibinfo{author}{\bibfnamefont{K.}~\bibnamefont{Tuominen}}, \bibnamefont{and}
  \bibinfo{author}{\bibfnamefont{M.}~\bibnamefont{Rosenlyst}},
  \bibinfo{journal}{Phys. Rev. D} \textbf{\bibinfo{volume}{108}},
  \bibinfo{pages}{015033} (\bibinfo{year}{2023}), \eprint{2301.00041}.

\bibitem[{\citenamefont{Hambye et~al.}(2018)\citenamefont{Hambye, Strumia, and
  Teresi}}]{Hambye:2018qjv}
\bibinfo{author}{\bibfnamefont{T.}~\bibnamefont{Hambye}},
  \bibinfo{author}{\bibfnamefont{A.}~\bibnamefont{Strumia}}, \bibnamefont{and}
  \bibinfo{author}{\bibfnamefont{D.}~\bibnamefont{Teresi}},
  \bibinfo{journal}{JHEP} \textbf{\bibinfo{volume}{08}}, \bibinfo{pages}{188}
  (\bibinfo{year}{2018}), \eprint{1805.01473}.

\bibitem[{\citenamefont{Baldes and Garcia-Cely}(2019)}]{Baldes:2018emh}
\bibinfo{author}{\bibfnamefont{I.}~\bibnamefont{Baldes}} \bibnamefont{and}
  \bibinfo{author}{\bibfnamefont{C.}~\bibnamefont{Garcia-Cely}},
  \bibinfo{journal}{JHEP} \textbf{\bibinfo{volume}{05}}, \bibinfo{pages}{190}
  (\bibinfo{year}{2019}), \eprint{1809.01198}.

\bibitem[{\citenamefont{Wong and Xie}(2023)}]{Wong:2023qon}
\bibinfo{author}{\bibfnamefont{X.-R.} \bibnamefont{Wong}} \bibnamefont{and}
  \bibinfo{author}{\bibfnamefont{K.-P.} \bibnamefont{Xie}},
  \bibinfo{journal}{Phys. Rev. D} \textbf{\bibinfo{volume}{108}},
  \bibinfo{pages}{055035} (\bibinfo{year}{2023}), \eprint{2304.00908}.

\bibitem[{\citenamefont{Gouttenoire}(2024)}]{Gouttenoire:2023pxh}
\bibinfo{author}{\bibfnamefont{Y.}~\bibnamefont{Gouttenoire}},
  \bibinfo{journal}{Phys. Lett. B} \textbf{\bibinfo{volume}{855}},
  \bibinfo{pages}{138800} (\bibinfo{year}{2024}), \eprint{2311.13640}.

\bibitem[{\citenamefont{Salvio}(2024)}]{Salvio:2023blb}
\bibinfo{author}{\bibfnamefont{A.}~\bibnamefont{Salvio}},
  \bibinfo{journal}{Phys. Lett. B} \textbf{\bibinfo{volume}{852}},
  \bibinfo{pages}{138639} (\bibinfo{year}{2024}), \eprint{2312.04628}.

\bibitem[{\citenamefont{Salvio}(2023)}]{Salvio:2023ynn}
\bibinfo{author}{\bibfnamefont{A.}~\bibnamefont{Salvio}},
  \bibinfo{journal}{JCAP} \textbf{\bibinfo{volume}{12}}, \bibinfo{pages}{046}
  (\bibinfo{year}{2023}), \eprint{2307.04694}.

\bibitem[{\citenamefont{Conaci et~al.}(2024)\citenamefont{Conaci, Delle~Rose,
  Dev, and Ghoshal}}]{Conaci:2024tlc}
\bibinfo{author}{\bibfnamefont{A.}~\bibnamefont{Conaci}},
  \bibinfo{author}{\bibfnamefont{L.}~\bibnamefont{Delle~Rose}},
  \bibinfo{author}{\bibfnamefont{P.~S.~B.} \bibnamefont{Dev}},
  \bibnamefont{and} \bibinfo{author}{\bibfnamefont{A.}~\bibnamefont{Ghoshal}}
  (\bibinfo{year}{2024}), \eprint{2401.09411}.

\bibitem[{\citenamefont{Banerjee et~al.}(2024)\citenamefont{Banerjee, Dey, and
  Khalil}}]{Banerjee:2024fam}
\bibinfo{author}{\bibfnamefont{I.~K.} \bibnamefont{Banerjee}},
  \bibinfo{author}{\bibfnamefont{U.~K.} \bibnamefont{Dey}}, \bibnamefont{and}
  \bibinfo{author}{\bibfnamefont{S.}~\bibnamefont{Khalil}}
  (\bibinfo{year}{2024}), \eprint{2406.12518}.

\bibitem[{\citenamefont{Gildener and Weinberg}(1976)}]{Gildener:1976ih}
\bibinfo{author}{\bibfnamefont{E.}~\bibnamefont{Gildener}} \bibnamefont{and}
  \bibinfo{author}{\bibfnamefont{S.}~\bibnamefont{Weinberg}},
  \bibinfo{journal}{Phys. Rev. D} \textbf{\bibinfo{volume}{13}},
  \bibinfo{pages}{3333} (\bibinfo{year}{1976}).

\bibitem[{\citenamefont{Chataignier et~al.}(2018)\citenamefont{Chataignier,
  Prokopec, Schmidt, and \'Swie\.zewska}}]{Chataignier:2018kay}
\bibinfo{author}{\bibfnamefont{L.}~\bibnamefont{Chataignier}},
  \bibinfo{author}{\bibfnamefont{T.}~\bibnamefont{Prokopec}},
  \bibinfo{author}{\bibfnamefont{M.~G.} \bibnamefont{Schmidt}},
  \bibnamefont{and}
  \bibinfo{author}{\bibfnamefont{B.}~\bibnamefont{\'Swie\.zewska}},
  \bibinfo{journal}{JHEP} \textbf{\bibinfo{volume}{08}}, \bibinfo{pages}{083}
  (\bibinfo{year}{2018}), \eprint{1805.09292}.

\bibitem[{\citenamefont{Lewis and Sullivan}(2017)}]{Lewis:2017dme}
\bibinfo{author}{\bibfnamefont{I.~M.} \bibnamefont{Lewis}} \bibnamefont{and}
  \bibinfo{author}{\bibfnamefont{M.}~\bibnamefont{Sullivan}},
  \bibinfo{journal}{Phys. Rev. D} \textbf{\bibinfo{volume}{96}},
  \bibinfo{pages}{035037} (\bibinfo{year}{2017}), \eprint{1701.08774}.

\bibitem[{\citenamefont{Liu and Xie}(2021)}]{Liu:2021jyc}
\bibinfo{author}{\bibfnamefont{W.}~\bibnamefont{Liu}} \bibnamefont{and}
  \bibinfo{author}{\bibfnamefont{K.-P.} \bibnamefont{Xie}},
  \bibinfo{journal}{JHEP} \textbf{\bibinfo{volume}{04}}, \bibinfo{pages}{015}
  (\bibinfo{year}{2021}), \eprint{2101.10469}.

\bibitem[{\citenamefont{Li and Xie}(2023)}]{Li:2023bxy}
\bibinfo{author}{\bibfnamefont{S.-P.} \bibnamefont{Li}} \bibnamefont{and}
  \bibinfo{author}{\bibfnamefont{K.-P.} \bibnamefont{Xie}},
  \bibinfo{journal}{Phys. Rev. D} \textbf{\bibinfo{volume}{108}},
  \bibinfo{pages}{055018} (\bibinfo{year}{2023}), \eprint{2307.01086}.

\bibitem[{\citenamefont{Gershtein et~al.}(2021)\citenamefont{Gershtein, Knapen,
  and Redigolo}}]{Gershtein:2020mwi}
\bibinfo{author}{\bibfnamefont{Y.}~\bibnamefont{Gershtein}},
  \bibinfo{author}{\bibfnamefont{S.}~\bibnamefont{Knapen}}, \bibnamefont{and}
  \bibinfo{author}{\bibfnamefont{D.}~\bibnamefont{Redigolo}},
  \bibinfo{journal}{Phys. Lett. B} \textbf{\bibinfo{volume}{823}},
  \bibinfo{pages}{136758} (\bibinfo{year}{2021}), \eprint{2012.07864}.

\bibitem[{\citenamefont{Djouadi}(2008)}]{Djouadi:2005gi}
\bibinfo{author}{\bibfnamefont{A.}~\bibnamefont{Djouadi}},
  \bibinfo{journal}{Phys. Rept.} \textbf{\bibinfo{volume}{457}},
  \bibinfo{pages}{1} (\bibinfo{year}{2008}), \eprint{hep-ph/0503172}.

\bibitem[{\citenamefont{Batell et~al.}(2022)\citenamefont{Batell, Blinov,
  Hearty, and McGehee}}]{Batell:2022dpx}
\bibinfo{author}{\bibfnamefont{B.}~\bibnamefont{Batell}},
  \bibinfo{author}{\bibfnamefont{N.}~\bibnamefont{Blinov}},
  \bibinfo{author}{\bibfnamefont{C.}~\bibnamefont{Hearty}}, \bibnamefont{and}
  \bibinfo{author}{\bibfnamefont{R.}~\bibnamefont{McGehee}}, in
  \emph{\bibinfo{booktitle}{{Snowmass 2021}}} (\bibinfo{year}{2022}),
  \eprint{2207.06905}.

\bibitem[{\citenamefont{Anastasiou et~al.}(2016)\citenamefont{Anastasiou, Duhr,
  Dulat, Furlan, Gehrmann, Herzog, Lazopoulos, and
  Mistlberger}}]{Anastasiou:2016hlm}
\bibinfo{author}{\bibfnamefont{C.}~\bibnamefont{Anastasiou}},
  \bibinfo{author}{\bibfnamefont{C.}~\bibnamefont{Duhr}},
  \bibinfo{author}{\bibfnamefont{F.}~\bibnamefont{Dulat}},
  \bibinfo{author}{\bibfnamefont{E.}~\bibnamefont{Furlan}},
  \bibinfo{author}{\bibfnamefont{T.}~\bibnamefont{Gehrmann}},
  \bibinfo{author}{\bibfnamefont{F.}~\bibnamefont{Herzog}},
  \bibinfo{author}{\bibfnamefont{A.}~\bibnamefont{Lazopoulos}},
  \bibnamefont{and}
  \bibinfo{author}{\bibfnamefont{B.}~\bibnamefont{Mistlberger}},
  \bibinfo{journal}{JHEP} \textbf{\bibinfo{volume}{09}}, \bibinfo{pages}{037}
  (\bibinfo{year}{2016}), \eprint{1605.05761}.

\bibitem[{\citenamefont{Sirunyan et~al.}(2018)}]{CMS:2018amk}
\bibinfo{author}{\bibfnamefont{A.~M.} \bibnamefont{Sirunyan}}
  \bibnamefont{et~al.} (\bibinfo{collaboration}{CMS}), \bibinfo{journal}{JHEP}
  \textbf{\bibinfo{volume}{06}}, \bibinfo{pages}{127} (\bibinfo{year}{2018}),
  \bibinfo{note}{[Erratum: JHEP 03, 128 (2019)]}, \eprint{1804.01939}.

\bibitem[{\citenamefont{Delahaye et~al.}(2019)\citenamefont{Delahaye, Diemoz,
  Long, Mansouli\'e, Pastrone, Rivkin, Schulte, Skrinsky, and
  Wulzer}}]{Delahaye:2019omf}
\bibinfo{author}{\bibfnamefont{J.~P.} \bibnamefont{Delahaye}},
  \bibinfo{author}{\bibfnamefont{M.}~\bibnamefont{Diemoz}},
  \bibinfo{author}{\bibfnamefont{K.}~\bibnamefont{Long}},
  \bibinfo{author}{\bibfnamefont{B.}~\bibnamefont{Mansouli\'e}},
  \bibinfo{author}{\bibfnamefont{N.}~\bibnamefont{Pastrone}},
  \bibinfo{author}{\bibfnamefont{L.}~\bibnamefont{Rivkin}},
  \bibinfo{author}{\bibfnamefont{D.}~\bibnamefont{Schulte}},
  \bibinfo{author}{\bibfnamefont{A.}~\bibnamefont{Skrinsky}}, \bibnamefont{and}
  \bibinfo{author}{\bibfnamefont{A.}~\bibnamefont{Wulzer}}
  (\bibinfo{year}{2019}), \eprint{1901.06150}.

\bibitem[{\citenamefont{Aime et~al.}(2022)}]{Aime:2022flm}
\bibinfo{author}{\bibfnamefont{C.}~\bibnamefont{Aime}} \bibnamefont{et~al.}
  (\bibinfo{year}{2022}), \eprint{2203.07256}.

\bibitem[{\citenamefont{Accettura et~al.}(2023)}]{Accettura:2023ked}
\bibinfo{author}{\bibfnamefont{C.}~\bibnamefont{Accettura}}
  \bibnamefont{et~al.}, \bibinfo{journal}{Eur. Phys. J. C}
  \textbf{\bibinfo{volume}{83}}, \bibinfo{pages}{864} (\bibinfo{year}{2023}),
  \bibinfo{note}{[Erratum: Eur.Phys.J.C 84, 36 (2024)]}, \eprint{2303.08533}.

\bibitem[{\citenamefont{Han et~al.}(2021)\citenamefont{Han, Ma, and
  Xie}}]{Han:2020uid}
\bibinfo{author}{\bibfnamefont{T.}~\bibnamefont{Han}},
  \bibinfo{author}{\bibfnamefont{Y.}~\bibnamefont{Ma}}, \bibnamefont{and}
  \bibinfo{author}{\bibfnamefont{K.}~\bibnamefont{Xie}},
  \bibinfo{journal}{Phys. Rev. D} \textbf{\bibinfo{volume}{103}},
  \bibinfo{pages}{L031301} (\bibinfo{year}{2021}), \eprint{2007.14300}.

\bibitem[{\citenamefont{Alloul et~al.}(2014)\citenamefont{Alloul, Christensen,
  Degrande, Duhr, and Fuks}}]{Alloul:2013bka}
\bibinfo{author}{\bibfnamefont{A.}~\bibnamefont{Alloul}},
  \bibinfo{author}{\bibfnamefont{N.~D.} \bibnamefont{Christensen}},
  \bibinfo{author}{\bibfnamefont{C.}~\bibnamefont{Degrande}},
  \bibinfo{author}{\bibfnamefont{C.}~\bibnamefont{Duhr}}, \bibnamefont{and}
  \bibinfo{author}{\bibfnamefont{B.}~\bibnamefont{Fuks}},
  \bibinfo{journal}{Comput. Phys. Commun.} \textbf{\bibinfo{volume}{185}},
  \bibinfo{pages}{2250} (\bibinfo{year}{2014}), \eprint{1310.1921}.

\bibitem[{\citenamefont{Alwall et~al.}(2014)\citenamefont{Alwall, Frederix,
  Frixione, Hirschi, Maltoni, Mattelaer, Shao, Stelzer, Torrielli, and
  Zaro}}]{Alwall:2014hca}
\bibinfo{author}{\bibfnamefont{J.}~\bibnamefont{Alwall}},
  \bibinfo{author}{\bibfnamefont{R.}~\bibnamefont{Frederix}},
  \bibinfo{author}{\bibfnamefont{S.}~\bibnamefont{Frixione}},
  \bibinfo{author}{\bibfnamefont{V.}~\bibnamefont{Hirschi}},
  \bibinfo{author}{\bibfnamefont{F.}~\bibnamefont{Maltoni}},
  \bibinfo{author}{\bibfnamefont{O.}~\bibnamefont{Mattelaer}},
  \bibinfo{author}{\bibfnamefont{H.~S.} \bibnamefont{Shao}},
  \bibinfo{author}{\bibfnamefont{T.}~\bibnamefont{Stelzer}},
  \bibinfo{author}{\bibfnamefont{P.}~\bibnamefont{Torrielli}},
  \bibnamefont{and} \bibinfo{author}{\bibfnamefont{M.}~\bibnamefont{Zaro}},
  \bibinfo{journal}{JHEP} \textbf{\bibinfo{volume}{07}}, \bibinfo{pages}{079}
  (\bibinfo{year}{2014}), \eprint{1405.0301}.

\bibitem[{\citenamefont{Linde}(1983)}]{Linde:1981zj}
\bibinfo{author}{\bibfnamefont{A.~D.} \bibnamefont{Linde}},
  \bibinfo{journal}{Nucl. Phys. B} \textbf{\bibinfo{volume}{216}},
  \bibinfo{pages}{421} (\bibinfo{year}{1983}), \bibinfo{note}{[Erratum:
  Nucl.Phys.B 223, 544 (1983)]}.

\bibitem[{\citenamefont{Guth and Tye}(1980)}]{Guth:1979bh}
\bibinfo{author}{\bibfnamefont{A.~H.} \bibnamefont{Guth}} \bibnamefont{and}
  \bibinfo{author}{\bibfnamefont{S.~H.~H.} \bibnamefont{Tye}},
  \bibinfo{journal}{Phys. Rev. Lett.} \textbf{\bibinfo{volume}{44}},
  \bibinfo{pages}{631} (\bibinfo{year}{1980}), \bibinfo{note}{[Erratum:
  Phys.Rev.Lett. 44, 963 (1980)]}.

\bibitem[{\citenamefont{Guth and Weinberg}(1981)}]{Guth:1981uk}
\bibinfo{author}{\bibfnamefont{A.~H.} \bibnamefont{Guth}} \bibnamefont{and}
  \bibinfo{author}{\bibfnamefont{E.~J.} \bibnamefont{Weinberg}},
  \bibinfo{journal}{Phys. Rev. D} \textbf{\bibinfo{volume}{23}},
  \bibinfo{pages}{876} (\bibinfo{year}{1981}).

\bibitem[{\citenamefont{Rintoul and Torquato}(1997)}]{rintoul1997precise}
\bibinfo{author}{\bibfnamefont{M.~D.} \bibnamefont{Rintoul}} \bibnamefont{and}
  \bibinfo{author}{\bibfnamefont{S.}~\bibnamefont{Torquato}},
  \bibinfo{journal}{Journal of physics a: mathematical and general}
  \textbf{\bibinfo{volume}{30}}, \bibinfo{pages}{L585} (\bibinfo{year}{1997}).

\bibitem[{\citenamefont{Witten}(1981)}]{Witten:1980ez}
\bibinfo{author}{\bibfnamefont{E.}~\bibnamefont{Witten}},
  \bibinfo{journal}{Nucl. Phys. B} \textbf{\bibinfo{volume}{177}},
  \bibinfo{pages}{477} (\bibinfo{year}{1981}).

\bibitem[{\citenamefont{Konstandin and Servant}(2011)}]{Konstandin:2011dr}
\bibinfo{author}{\bibfnamefont{T.}~\bibnamefont{Konstandin}} \bibnamefont{and}
  \bibinfo{author}{\bibfnamefont{G.}~\bibnamefont{Servant}},
  \bibinfo{journal}{JCAP} \textbf{\bibinfo{volume}{12}}, \bibinfo{pages}{009}
  (\bibinfo{year}{2011}), \eprint{1104.4791}.

\bibitem[{\citenamefont{Jinno and Takimoto}(2017)}]{Jinno:2016knw}
\bibinfo{author}{\bibfnamefont{R.}~\bibnamefont{Jinno}} \bibnamefont{and}
  \bibinfo{author}{\bibfnamefont{M.}~\bibnamefont{Takimoto}},
  \bibinfo{journal}{Phys. Rev. D} \textbf{\bibinfo{volume}{95}},
  \bibinfo{pages}{015020} (\bibinfo{year}{2017}), \eprint{1604.05035}.

\bibitem[{\citenamefont{Iso et~al.}(2017)\citenamefont{Iso, Serpico, and
  Shimada}}]{Iso:2017uuu}
\bibinfo{author}{\bibfnamefont{S.}~\bibnamefont{Iso}},
  \bibinfo{author}{\bibfnamefont{P.~D.} \bibnamefont{Serpico}},
  \bibnamefont{and} \bibinfo{author}{\bibfnamefont{K.}~\bibnamefont{Shimada}},
  \bibinfo{journal}{Phys. Rev. Lett.} \textbf{\bibinfo{volume}{119}},
  \bibinfo{pages}{141301} (\bibinfo{year}{2017}), \eprint{1704.04955}.

\bibitem[{\citenamefont{Ghorbani}(2018)}]{Ghorbani:2017lyk}
\bibinfo{author}{\bibfnamefont{P.~H.} \bibnamefont{Ghorbani}},
  \bibinfo{journal}{Phys. Rev. D} \textbf{\bibinfo{volume}{98}},
  \bibinfo{pages}{115016} (\bibinfo{year}{2018}), \eprint{1711.11541}.

\bibitem[{\citenamefont{Marzo et~al.}(2019)\citenamefont{Marzo, Marzola, and
  Vaskonen}}]{Marzo:2018nov}
\bibinfo{author}{\bibfnamefont{C.}~\bibnamefont{Marzo}},
  \bibinfo{author}{\bibfnamefont{L.}~\bibnamefont{Marzola}}, \bibnamefont{and}
  \bibinfo{author}{\bibfnamefont{V.}~\bibnamefont{Vaskonen}},
  \bibinfo{journal}{Eur. Phys. J. C} \textbf{\bibinfo{volume}{79}},
  \bibinfo{pages}{601} (\bibinfo{year}{2019}), \eprint{1811.11169}.

\bibitem[{\citenamefont{Bian et~al.}(2021)\citenamefont{Bian, Cheng, Guo, and
  Zhang}}]{Bian:2019szo}
\bibinfo{author}{\bibfnamefont{L.}~\bibnamefont{Bian}},
  \bibinfo{author}{\bibfnamefont{W.}~\bibnamefont{Cheng}},
  \bibinfo{author}{\bibfnamefont{H.-K.} \bibnamefont{Guo}}, \bibnamefont{and}
  \bibinfo{author}{\bibfnamefont{Y.}~\bibnamefont{Zhang}},
  \bibinfo{journal}{Chin. Phys. C} \textbf{\bibinfo{volume}{45}},
  \bibinfo{pages}{113104} (\bibinfo{year}{2021}), \eprint{1907.13589}.

\bibitem[{\citenamefont{Ellis et~al.}(2019{\natexlab{a}})\citenamefont{Ellis,
  Lewicki, No, and Vaskonen}}]{Ellis:2019oqb}
\bibinfo{author}{\bibfnamefont{J.}~\bibnamefont{Ellis}},
  \bibinfo{author}{\bibfnamefont{M.}~\bibnamefont{Lewicki}},
  \bibinfo{author}{\bibfnamefont{J.~M.} \bibnamefont{No}}, \bibnamefont{and}
  \bibinfo{author}{\bibfnamefont{V.}~\bibnamefont{Vaskonen}},
  \bibinfo{journal}{JCAP} \textbf{\bibinfo{volume}{06}}, \bibinfo{pages}{024}
  (\bibinfo{year}{2019}{\natexlab{a}}), \eprint{1903.09642}.

\bibitem[{\citenamefont{Ellis et~al.}(2020)\citenamefont{Ellis, Lewicki, and
  Vaskonen}}]{Ellis:2020nnr}
\bibinfo{author}{\bibfnamefont{J.}~\bibnamefont{Ellis}},
  \bibinfo{author}{\bibfnamefont{M.}~\bibnamefont{Lewicki}}, \bibnamefont{and}
  \bibinfo{author}{\bibfnamefont{V.}~\bibnamefont{Vaskonen}},
  \bibinfo{journal}{JCAP} \textbf{\bibinfo{volume}{11}}, \bibinfo{pages}{020}
  (\bibinfo{year}{2020}), \eprint{2007.15586}.

\bibitem[{\citenamefont{Jung and Kawana}(2022)}]{Jung:2021vap}
\bibinfo{author}{\bibfnamefont{S.}~\bibnamefont{Jung}} \bibnamefont{and}
  \bibinfo{author}{\bibfnamefont{K.}~\bibnamefont{Kawana}},
  \bibinfo{journal}{PTEP} \textbf{\bibinfo{volume}{2022}},
  \bibinfo{pages}{033B11} (\bibinfo{year}{2022}), \eprint{2105.01217}.

\bibitem[{\citenamefont{Kawana}(2022)}]{Kawana:2022fum}
\bibinfo{author}{\bibfnamefont{K.}~\bibnamefont{Kawana}},
  \bibinfo{journal}{Phys. Rev. D} \textbf{\bibinfo{volume}{105}},
  \bibinfo{pages}{103515} (\bibinfo{year}{2022}), \eprint{2201.00560}.

\bibitem[{\citenamefont{Zhao et~al.}(2023)\citenamefont{Zhao, Di, Bian, and
  Cai}}]{Zhao:2022cnn}
\bibinfo{author}{\bibfnamefont{Z.}~\bibnamefont{Zhao}},
  \bibinfo{author}{\bibfnamefont{Y.}~\bibnamefont{Di}},
  \bibinfo{author}{\bibfnamefont{L.}~\bibnamefont{Bian}}, \bibnamefont{and}
  \bibinfo{author}{\bibfnamefont{R.-G.} \bibnamefont{Cai}},
  \bibinfo{journal}{JHEP} \textbf{\bibinfo{volume}{10}}, \bibinfo{pages}{158}
  (\bibinfo{year}{2023}), \eprint{2204.04427}.

\bibitem[{\citenamefont{Sagunski et~al.}(2023)\citenamefont{Sagunski, Schicho,
  and Schmitt}}]{Sagunski:2023ynd}
\bibinfo{author}{\bibfnamefont{L.}~\bibnamefont{Sagunski}},
  \bibinfo{author}{\bibfnamefont{P.}~\bibnamefont{Schicho}}, \bibnamefont{and}
  \bibinfo{author}{\bibfnamefont{D.}~\bibnamefont{Schmitt}},
  \bibinfo{journal}{Phys. Rev. D} \textbf{\bibinfo{volume}{107}},
  \bibinfo{pages}{123512} (\bibinfo{year}{2023}), \eprint{2303.02450}.

\bibitem[{\citenamefont{Ahriche et~al.}(2024)\citenamefont{Ahriche, Kanemura,
  and Tanaka}}]{Ahriche:2023jdq}
\bibinfo{author}{\bibfnamefont{A.}~\bibnamefont{Ahriche}},
  \bibinfo{author}{\bibfnamefont{S.}~\bibnamefont{Kanemura}}, \bibnamefont{and}
  \bibinfo{author}{\bibfnamefont{M.}~\bibnamefont{Tanaka}},
  \bibinfo{journal}{JHEP} \textbf{\bibinfo{volume}{01}}, \bibinfo{pages}{201}
  (\bibinfo{year}{2024}), \eprint{2308.12676}.

\bibitem[{\citenamefont{Ghorbani}(2024)}]{Ghorbani:2024twk}
\bibinfo{author}{\bibfnamefont{P.~H.} \bibnamefont{Ghorbani}}
  (\bibinfo{year}{2024}), \eprint{2408.16475}.

\bibitem[{\citenamefont{Braun and Gies}(2006)}]{Braun:2006jd}
\bibinfo{author}{\bibfnamefont{J.}~\bibnamefont{Braun}} \bibnamefont{and}
  \bibinfo{author}{\bibfnamefont{H.}~\bibnamefont{Gies}},
  \bibinfo{journal}{JHEP} \textbf{\bibinfo{volume}{06}}, \bibinfo{pages}{024}
  (\bibinfo{year}{2006}), \eprint{hep-ph/0602226}.

\bibitem[{\citenamefont{Pisarski and Wilczek}(1984)}]{Pisarski:1983ms}
\bibinfo{author}{\bibfnamefont{R.~D.} \bibnamefont{Pisarski}} \bibnamefont{and}
  \bibinfo{author}{\bibfnamefont{F.}~\bibnamefont{Wilczek}},
  \bibinfo{journal}{Phys. Rev. D} \textbf{\bibinfo{volume}{29}},
  \bibinfo{pages}{338} (\bibinfo{year}{1984}).

\bibitem[{\citenamefont{Guan and Matsuzaki}(2024)}]{Guan:2024ccw}
\bibinfo{author}{\bibfnamefont{Y.}~\bibnamefont{Guan}} \bibnamefont{and}
  \bibinfo{author}{\bibfnamefont{S.}~\bibnamefont{Matsuzaki}}
  (\bibinfo{year}{2024}), \eprint{2405.03265}.

\bibitem[{\citenamefont{Ellis et~al.}(2019{\natexlab{b}})\citenamefont{Ellis,
  Lewicki, and No}}]{Ellis:2018mja}
\bibinfo{author}{\bibfnamefont{J.}~\bibnamefont{Ellis}},
  \bibinfo{author}{\bibfnamefont{M.}~\bibnamefont{Lewicki}}, \bibnamefont{and}
  \bibinfo{author}{\bibfnamefont{J.~M.} \bibnamefont{No}},
  \bibinfo{journal}{JCAP} \textbf{\bibinfo{volume}{04}}, \bibinfo{pages}{003}
  (\bibinfo{year}{2019}{\natexlab{b}}), \eprint{1809.08242}.

\bibitem[{\citenamefont{Turner et~al.}(1992)\citenamefont{Turner, Weinberg, and
  Widrow}}]{Turner:1992tz}
\bibinfo{author}{\bibfnamefont{M.~S.} \bibnamefont{Turner}},
  \bibinfo{author}{\bibfnamefont{E.~J.} \bibnamefont{Weinberg}},
  \bibnamefont{and} \bibinfo{author}{\bibfnamefont{L.~M.}
  \bibnamefont{Widrow}}, \bibinfo{journal}{Phys. Rev. D}
  \textbf{\bibinfo{volume}{46}}, \bibinfo{pages}{2384} (\bibinfo{year}{1992}).

\bibitem[{\citenamefont{Lyth and Stewart}(1996)}]{Lyth:1995ka}
\bibinfo{author}{\bibfnamefont{D.~H.} \bibnamefont{Lyth}} \bibnamefont{and}
  \bibinfo{author}{\bibfnamefont{E.~D.} \bibnamefont{Stewart}},
  \bibinfo{journal}{Phys. Rev. D} \textbf{\bibinfo{volume}{53}},
  \bibinfo{pages}{1784} (\bibinfo{year}{1996}), \eprint{hep-ph/9510204}.

\bibitem[{\citenamefont{Dutra and Wu}(2023)}]{Dutra:2021phm}
\bibinfo{author}{\bibfnamefont{M.}~\bibnamefont{Dutra}} \bibnamefont{and}
  \bibinfo{author}{\bibfnamefont{Y.}~\bibnamefont{Wu}}, \bibinfo{journal}{Phys.
  Dark Univ.} \textbf{\bibinfo{volume}{40}}, \bibinfo{pages}{101198}
  (\bibinfo{year}{2023}), \eprint{2111.15665}.

\bibitem[{\citenamefont{Espinosa et~al.}(2010)\citenamefont{Espinosa,
  Konstandin, No, and Servant}}]{Espinosa:2010hh}
\bibinfo{author}{\bibfnamefont{J.~R.} \bibnamefont{Espinosa}},
  \bibinfo{author}{\bibfnamefont{T.}~\bibnamefont{Konstandin}},
  \bibinfo{author}{\bibfnamefont{J.~M.} \bibnamefont{No}}, \bibnamefont{and}
  \bibinfo{author}{\bibfnamefont{G.}~\bibnamefont{Servant}},
  \bibinfo{journal}{JCAP} \textbf{\bibinfo{volume}{06}}, \bibinfo{pages}{028}
  (\bibinfo{year}{2010}), \eprint{1004.4187}.

\bibitem[{\citenamefont{Giese et~al.}(2020)\citenamefont{Giese, Konstandin, and
  van~de Vis}}]{Giese:2020rtr}
\bibinfo{author}{\bibfnamefont{F.}~\bibnamefont{Giese}},
  \bibinfo{author}{\bibfnamefont{T.}~\bibnamefont{Konstandin}},
  \bibnamefont{and} \bibinfo{author}{\bibfnamefont{J.}~\bibnamefont{van~de
  Vis}}, \bibinfo{journal}{JCAP} \textbf{\bibinfo{volume}{07}},
  \bibinfo{pages}{057} (\bibinfo{year}{2020}), \eprint{2004.06995}.

\bibitem[{\citenamefont{Wang and Yuwen}(2022)}]{Wang:2022lyd}
\bibinfo{author}{\bibfnamefont{S.-J.} \bibnamefont{Wang}} \bibnamefont{and}
  \bibinfo{author}{\bibfnamefont{Z.-Y.} \bibnamefont{Yuwen}},
  \bibinfo{journal}{JCAP} \textbf{\bibinfo{volume}{10}}, \bibinfo{pages}{047}
  (\bibinfo{year}{2022}), \eprint{2206.01148}.

\bibitem[{\citenamefont{Roshan and White}(2024)}]{Roshan:2024qnv}
\bibinfo{author}{\bibfnamefont{R.}~\bibnamefont{Roshan}} \bibnamefont{and}
  \bibinfo{author}{\bibfnamefont{G.}~\bibnamefont{White}}
  (\bibinfo{year}{2024}), \eprint{2401.04388}.

\bibitem[{\citenamefont{Bodeker and Moore}(2017)}]{Bodeker:2017cim}
\bibinfo{author}{\bibfnamefont{D.}~\bibnamefont{Bodeker}} \bibnamefont{and}
  \bibinfo{author}{\bibfnamefont{G.~D.} \bibnamefont{Moore}},
  \bibinfo{journal}{JCAP} \textbf{\bibinfo{volume}{05}}, \bibinfo{pages}{025}
  (\bibinfo{year}{2017}), \eprint{1703.08215}.

\bibitem[{\citenamefont{Gouttenoire et~al.}(2022)\citenamefont{Gouttenoire,
  Jinno, and Sala}}]{Gouttenoire:2021kjv}
\bibinfo{author}{\bibfnamefont{Y.}~\bibnamefont{Gouttenoire}},
  \bibinfo{author}{\bibfnamefont{R.}~\bibnamefont{Jinno}}, \bibnamefont{and}
  \bibinfo{author}{\bibfnamefont{F.}~\bibnamefont{Sala}},
  \bibinfo{journal}{JHEP} \textbf{\bibinfo{volume}{05}}, \bibinfo{pages}{004}
  (\bibinfo{year}{2022}), \eprint{2112.07686}.

\bibitem[{\citenamefont{H\"oche et~al.}(2021)\citenamefont{H\"oche, Kozaczuk,
  Long, Turner, and Wang}}]{Hoche:2020ysm}
\bibinfo{author}{\bibfnamefont{S.}~\bibnamefont{H\"oche}},
  \bibinfo{author}{\bibfnamefont{J.}~\bibnamefont{Kozaczuk}},
  \bibinfo{author}{\bibfnamefont{A.~J.} \bibnamefont{Long}},
  \bibinfo{author}{\bibfnamefont{J.}~\bibnamefont{Turner}}, \bibnamefont{and}
  \bibinfo{author}{\bibfnamefont{Y.}~\bibnamefont{Wang}},
  \bibinfo{journal}{JCAP} \textbf{\bibinfo{volume}{03}}, \bibinfo{pages}{009}
  (\bibinfo{year}{2021}), \eprint{2007.10343}.

\bibitem[{\citenamefont{Kanemura et~al.}(2024)\citenamefont{Kanemura, Tanaka,
  and Xie}}]{Kanemura:2024pae}
\bibinfo{author}{\bibfnamefont{S.}~\bibnamefont{Kanemura}},
  \bibinfo{author}{\bibfnamefont{M.}~\bibnamefont{Tanaka}}, \bibnamefont{and}
  \bibinfo{author}{\bibfnamefont{K.-P.} \bibnamefont{Xie}},
  \bibinfo{journal}{JHEP} \textbf{\bibinfo{volume}{06}}, \bibinfo{pages}{036}
  (\bibinfo{year}{2024}), \eprint{2404.00646}.

\bibitem[{\citenamefont{Wang et~al.}(2020)\citenamefont{Wang, Huang, and
  Zhang}}]{Wang:2020jrd}
\bibinfo{author}{\bibfnamefont{X.}~\bibnamefont{Wang}},
  \bibinfo{author}{\bibfnamefont{F.~P.} \bibnamefont{Huang}}, \bibnamefont{and}
  \bibinfo{author}{\bibfnamefont{X.}~\bibnamefont{Zhang}},
  \bibinfo{journal}{JCAP} \textbf{\bibinfo{volume}{05}}, \bibinfo{pages}{045}
  (\bibinfo{year}{2020}), \eprint{2003.08892}.

\bibitem[{\citenamefont{Huber and Konstandin}(2008)}]{Huber:2008hg}
\bibinfo{author}{\bibfnamefont{S.~J.} \bibnamefont{Huber}} \bibnamefont{and}
  \bibinfo{author}{\bibfnamefont{T.}~\bibnamefont{Konstandin}},
  \bibinfo{journal}{JCAP} \textbf{\bibinfo{volume}{09}}, \bibinfo{pages}{022}
  (\bibinfo{year}{2008}), \eprint{0806.1828}.

\bibitem[{\citenamefont{Hindmarsh et~al.}(2015)\citenamefont{Hindmarsh, Huber,
  Rummukainen, and Weir}}]{Hindmarsh:2015qta}
\bibinfo{author}{\bibfnamefont{M.}~\bibnamefont{Hindmarsh}},
  \bibinfo{author}{\bibfnamefont{S.~J.} \bibnamefont{Huber}},
  \bibinfo{author}{\bibfnamefont{K.}~\bibnamefont{Rummukainen}},
  \bibnamefont{and} \bibinfo{author}{\bibfnamefont{D.~J.} \bibnamefont{Weir}},
  \bibinfo{journal}{Phys. Rev. D} \textbf{\bibinfo{volume}{92}},
  \bibinfo{pages}{123009} (\bibinfo{year}{2015}), \eprint{1504.03291}.

\bibitem[{\citenamefont{Caprini et~al.}(2009)\citenamefont{Caprini, Durrer, and
  Servant}}]{Caprini:2009yp}
\bibinfo{author}{\bibfnamefont{C.}~\bibnamefont{Caprini}},
  \bibinfo{author}{\bibfnamefont{R.}~\bibnamefont{Durrer}}, \bibnamefont{and}
  \bibinfo{author}{\bibfnamefont{G.}~\bibnamefont{Servant}},
  \bibinfo{journal}{JCAP} \textbf{\bibinfo{volume}{12}}, \bibinfo{pages}{024}
  (\bibinfo{year}{2009}), \eprint{0909.0622}.

\bibitem[{\citenamefont{Lewicki and Vaskonen}(2023)}]{Lewicki:2022pdb}
\bibinfo{author}{\bibfnamefont{M.}~\bibnamefont{Lewicki}} \bibnamefont{and}
  \bibinfo{author}{\bibfnamefont{V.}~\bibnamefont{Vaskonen}},
  \bibinfo{journal}{Eur. Phys. J. C} \textbf{\bibinfo{volume}{83}},
  \bibinfo{pages}{109} (\bibinfo{year}{2023}), \eprint{2208.11697}.

\bibitem[{\citenamefont{Ellis et~al.}(2024)\citenamefont{Ellis, Fairbairn,
  Franciolini, H\"utsi, Iovino, Lewicki, Raidal, Urrutia, Vaskonen, and
  Veerm\"ae}}]{Ellis:2023oxs}
\bibinfo{author}{\bibfnamefont{J.}~\bibnamefont{Ellis}},
  \bibinfo{author}{\bibfnamefont{M.}~\bibnamefont{Fairbairn}},
  \bibinfo{author}{\bibfnamefont{G.}~\bibnamefont{Franciolini}},
  \bibinfo{author}{\bibfnamefont{G.}~\bibnamefont{H\"utsi}},
  \bibinfo{author}{\bibfnamefont{A.}~\bibnamefont{Iovino}},
  \bibinfo{author}{\bibfnamefont{M.}~\bibnamefont{Lewicki}},
  \bibinfo{author}{\bibfnamefont{M.}~\bibnamefont{Raidal}},
  \bibinfo{author}{\bibfnamefont{J.}~\bibnamefont{Urrutia}},
  \bibinfo{author}{\bibfnamefont{V.}~\bibnamefont{Vaskonen}}, \bibnamefont{and}
  \bibinfo{author}{\bibfnamefont{H.}~\bibnamefont{Veerm\"ae}},
  \bibinfo{journal}{Phys. Rev. D} \textbf{\bibinfo{volume}{109}},
  \bibinfo{pages}{023522} (\bibinfo{year}{2024}), \eprint{2308.08546}.

\bibitem[{\citenamefont{Caprini et~al.}(2024)\citenamefont{Caprini, Jinno,
  Lewicki, Madge, Merchand, Nardini, Pieroni, Roper~Pol, and
  Vaskonen}}]{Caprini:2024hue}
\bibinfo{author}{\bibfnamefont{C.}~\bibnamefont{Caprini}},
  \bibinfo{author}{\bibfnamefont{R.}~\bibnamefont{Jinno}},
  \bibinfo{author}{\bibfnamefont{M.}~\bibnamefont{Lewicki}},
  \bibinfo{author}{\bibfnamefont{E.}~\bibnamefont{Madge}},
  \bibinfo{author}{\bibfnamefont{M.}~\bibnamefont{Merchand}},
  \bibinfo{author}{\bibfnamefont{G.}~\bibnamefont{Nardini}},
  \bibinfo{author}{\bibfnamefont{M.}~\bibnamefont{Pieroni}},
  \bibinfo{author}{\bibfnamefont{A.}~\bibnamefont{Roper~Pol}},
  \bibnamefont{and} \bibinfo{author}{\bibfnamefont{V.}~\bibnamefont{Vaskonen}}
  (\bibinfo{collaboration}{LISA Cosmology Working Group})
  (\bibinfo{year}{2024}), \eprint{2403.03723}.

\bibitem[{\citenamefont{Breitbach et~al.}(2019)\citenamefont{Breitbach, Kopp,
  Madge, Opferkuch, and Schwaller}}]{Breitbach:2018ddu}
\bibinfo{author}{\bibfnamefont{M.}~\bibnamefont{Breitbach}},
  \bibinfo{author}{\bibfnamefont{J.}~\bibnamefont{Kopp}},
  \bibinfo{author}{\bibfnamefont{E.}~\bibnamefont{Madge}},
  \bibinfo{author}{\bibfnamefont{T.}~\bibnamefont{Opferkuch}},
  \bibnamefont{and}
  \bibinfo{author}{\bibfnamefont{P.}~\bibnamefont{Schwaller}},
  \bibinfo{journal}{JCAP} \textbf{\bibinfo{volume}{07}}, \bibinfo{pages}{007}
  (\bibinfo{year}{2019}), \eprint{1811.11175}.

\bibitem[{\citenamefont{Caprini et~al.}(2016)}]{Caprini:2015zlo}
\bibinfo{author}{\bibfnamefont{C.}~\bibnamefont{Caprini}} \bibnamefont{et~al.},
  \bibinfo{journal}{JCAP} \textbf{\bibinfo{volume}{04}}, \bibinfo{pages}{001}
  (\bibinfo{year}{2016}), \eprint{1512.06239}.

\bibitem[{\citenamefont{Caprini et~al.}(2020)}]{Caprini:2019egz}
\bibinfo{author}{\bibfnamefont{C.}~\bibnamefont{Caprini}} \bibnamefont{et~al.},
  \bibinfo{journal}{JCAP} \textbf{\bibinfo{volume}{03}}, \bibinfo{pages}{024}
  (\bibinfo{year}{2020}), \eprint{1910.13125}.

\bibitem[{\citenamefont{Amaro-Seoane et~al.}(2017)}]{LISA:2017pwj}
\bibinfo{author}{\bibfnamefont{P.}~\bibnamefont{Amaro-Seoane}}
  \bibnamefont{et~al.} (\bibinfo{collaboration}{LISA}) (\bibinfo{year}{2017}),
  \eprint{1702.00786}.

\bibitem[{\citenamefont{Luo et~al.}(2016)}]{TianQin:2015yph}
\bibinfo{author}{\bibfnamefont{J.}~\bibnamefont{Luo}} \bibnamefont{et~al.}
  (\bibinfo{collaboration}{TianQin}), \bibinfo{journal}{Class. Quant. Grav.}
  \textbf{\bibinfo{volume}{33}}, \bibinfo{pages}{035010}
  (\bibinfo{year}{2016}), \eprint{1512.02076}.

\bibitem[{\citenamefont{Hu and Wu}(2017)}]{Hu:2017mde}
\bibinfo{author}{\bibfnamefont{W.-R.} \bibnamefont{Hu}} \bibnamefont{and}
  \bibinfo{author}{\bibfnamefont{Y.-L.} \bibnamefont{Wu}},
  \bibinfo{journal}{Natl. Sci. Rev.} \textbf{\bibinfo{volume}{4}},
  \bibinfo{pages}{685} (\bibinfo{year}{2017}).

\bibitem[{\citenamefont{Crowder and Cornish}(2005)}]{Crowder:2005nr}
\bibinfo{author}{\bibfnamefont{J.}~\bibnamefont{Crowder}} \bibnamefont{and}
  \bibinfo{author}{\bibfnamefont{N.~J.} \bibnamefont{Cornish}},
  \bibinfo{journal}{Phys. Rev. D} \textbf{\bibinfo{volume}{72}},
  \bibinfo{pages}{083005} (\bibinfo{year}{2005}), \eprint{gr-qc/0506015}.

\bibitem[{\citenamefont{Aad et~al.}(2024)}]{ATLAS:2024fkg}
\bibinfo{author}{\bibfnamefont{G.}~\bibnamefont{Aad}} \bibnamefont{et~al.}
  (\bibinfo{collaboration}{ATLAS}) (\bibinfo{year}{2024}), \eprint{2404.05498}.

\bibitem[{\citenamefont{Schmitz}(2021)}]{Schmitz:2020syl}
\bibinfo{author}{\bibfnamefont{K.}~\bibnamefont{Schmitz}},
  \bibinfo{journal}{JHEP} \textbf{\bibinfo{volume}{01}}, \bibinfo{pages}{097}
  (\bibinfo{year}{2021}), \eprint{2002.04615}.

\bibitem[{\citenamefont{Athron et~al.}(2024)\citenamefont{Athron, Morris, and
  Xu}}]{Athron:2023rfq}
\bibinfo{author}{\bibfnamefont{P.}~\bibnamefont{Athron}},
  \bibinfo{author}{\bibfnamefont{L.}~\bibnamefont{Morris}}, \bibnamefont{and}
  \bibinfo{author}{\bibfnamefont{Z.}~\bibnamefont{Xu}}, \bibinfo{journal}{JCAP}
  \textbf{\bibinfo{volume}{05}}, \bibinfo{pages}{075} (\bibinfo{year}{2024}),
  \eprint{2309.05474}.

\bibitem[{\citenamefont{Guo et~al.}(2021)\citenamefont{Guo, Sinha, Vagie, and
  White}}]{Guo:2021qcq}
\bibinfo{author}{\bibfnamefont{H.-K.} \bibnamefont{Guo}},
  \bibinfo{author}{\bibfnamefont{K.}~\bibnamefont{Sinha}},
  \bibinfo{author}{\bibfnamefont{D.}~\bibnamefont{Vagie}}, \bibnamefont{and}
  \bibinfo{author}{\bibfnamefont{G.}~\bibnamefont{White}},
  \bibinfo{journal}{JHEP} \textbf{\bibinfo{volume}{06}}, \bibinfo{pages}{164}
  (\bibinfo{year}{2021}), \eprint{2103.06933}.

\bibitem[{\citenamefont{Alves et~al.}(2018)\citenamefont{Alves, Ghosh, Guo, and
  Sinha}}]{Alves:2018oct}
\bibinfo{author}{\bibfnamefont{A.}~\bibnamefont{Alves}},
  \bibinfo{author}{\bibfnamefont{T.}~\bibnamefont{Ghosh}},
  \bibinfo{author}{\bibfnamefont{H.-K.} \bibnamefont{Guo}}, \bibnamefont{and}
  \bibinfo{author}{\bibfnamefont{K.}~\bibnamefont{Sinha}},
  \bibinfo{journal}{JHEP} \textbf{\bibinfo{volume}{12}}, \bibinfo{pages}{070}
  (\bibinfo{year}{2018}), \eprint{1808.08974}.

\bibitem[{\citenamefont{Alves et~al.}(2020)\citenamefont{Alves, Gon\c{c}alves,
  Ghosh, Guo, and Sinha}}]{Alves:2019igs}
\bibinfo{author}{\bibfnamefont{A.}~\bibnamefont{Alves}},
  \bibinfo{author}{\bibfnamefont{D.}~\bibnamefont{Gon\c{c}alves}},
  \bibinfo{author}{\bibfnamefont{T.}~\bibnamefont{Ghosh}},
  \bibinfo{author}{\bibfnamefont{H.-K.} \bibnamefont{Guo}}, \bibnamefont{and}
  \bibinfo{author}{\bibfnamefont{K.}~\bibnamefont{Sinha}},
  \bibinfo{journal}{JHEP} \textbf{\bibinfo{volume}{03}}, \bibinfo{pages}{053}
  (\bibinfo{year}{2020}), \eprint{1909.05268}.

\bibitem[{\citenamefont{Alves et~al.}(2021)\citenamefont{Alves, Gon\c{c}alves,
  Ghosh, Guo, and Sinha}}]{Alves:2020bpi}
\bibinfo{author}{\bibfnamefont{A.}~\bibnamefont{Alves}},
  \bibinfo{author}{\bibfnamefont{D.}~\bibnamefont{Gon\c{c}alves}},
  \bibinfo{author}{\bibfnamefont{T.}~\bibnamefont{Ghosh}},
  \bibinfo{author}{\bibfnamefont{H.-K.} \bibnamefont{Guo}}, \bibnamefont{and}
  \bibinfo{author}{\bibfnamefont{K.}~\bibnamefont{Sinha}},
  \bibinfo{journal}{Phys. Lett. B} \textbf{\bibinfo{volume}{818}},
  \bibinfo{pages}{136377} (\bibinfo{year}{2021}), \eprint{2007.15654}.

\bibitem[{\citenamefont{Aaij et~al.}(2017)}]{LHCb:2016awg}
\bibinfo{author}{\bibfnamefont{R.}~\bibnamefont{Aaij}} \bibnamefont{et~al.}
  (\bibinfo{collaboration}{LHCb}), \bibinfo{journal}{Phys. Rev. D}
  \textbf{\bibinfo{volume}{95}}, \bibinfo{pages}{071101}
  (\bibinfo{year}{2017}), \eprint{1612.07818}.

\bibitem[{\citenamefont{Aaij et~al.}(2015)}]{LHCb:2015nkv}
\bibinfo{author}{\bibfnamefont{R.}~\bibnamefont{Aaij}} \bibnamefont{et~al.}
  (\bibinfo{collaboration}{LHCb}), \bibinfo{journal}{Phys. Rev. Lett.}
  \textbf{\bibinfo{volume}{115}}, \bibinfo{pages}{161802}
  (\bibinfo{year}{2015}), \eprint{1508.04094}.

\bibitem[{\citenamefont{Cortina~Gil et~al.}(2021{\natexlab{a}})}]{NA62:2020pwi}
\bibinfo{author}{\bibfnamefont{E.}~\bibnamefont{Cortina~Gil}}
  \bibnamefont{et~al.} (\bibinfo{collaboration}{NA62}), \bibinfo{journal}{JHEP}
  \textbf{\bibinfo{volume}{02}}, \bibinfo{pages}{201}
  (\bibinfo{year}{2021}{\natexlab{a}}), \eprint{2010.07644}.

\bibitem[{\citenamefont{Cortina~Gil et~al.}(2021{\natexlab{b}})}]{NA62:2020xlg}
\bibinfo{author}{\bibfnamefont{E.}~\bibnamefont{Cortina~Gil}}
  \bibnamefont{et~al.} (\bibinfo{collaboration}{NA62}), \bibinfo{journal}{JHEP}
  \textbf{\bibinfo{volume}{03}}, \bibinfo{pages}{058}
  (\bibinfo{year}{2021}{\natexlab{b}}), \eprint{2011.11329}.

\bibitem[{\citenamefont{Winkler}(2019)}]{Winkler:2018qyg}
\bibinfo{author}{\bibfnamefont{M.~W.} \bibnamefont{Winkler}},
  \bibinfo{journal}{Phys. Rev. D} \textbf{\bibinfo{volume}{99}},
  \bibinfo{pages}{015018} (\bibinfo{year}{2019}), \eprint{1809.01876}.

\bibitem[{\citenamefont{Artamonov et~al.}(2009)}]{BNL-E949:2009dza}
\bibinfo{author}{\bibfnamefont{A.~V.} \bibnamefont{Artamonov}}
  \bibnamefont{et~al.} (\bibinfo{collaboration}{BNL-E949}),
  \bibinfo{journal}{Phys. Rev. D} \textbf{\bibinfo{volume}{79}},
  \bibinfo{pages}{092004} (\bibinfo{year}{2009}), \eprint{0903.0030}.

\bibitem[{\citenamefont{Foroughi-Abari and
  Ritz}(2020)}]{Foroughi-Abari:2020gju}
\bibinfo{author}{\bibfnamefont{S.}~\bibnamefont{Foroughi-Abari}}
  \bibnamefont{and} \bibinfo{author}{\bibfnamefont{A.}~\bibnamefont{Ritz}},
  \bibinfo{journal}{Phys. Rev. D} \textbf{\bibinfo{volume}{102}},
  \bibinfo{pages}{035015} (\bibinfo{year}{2020}), \eprint{2004.14515}.

\bibitem[{\citenamefont{Abratenko et~al.}(2021)}]{MicroBooNE:2020vlq}
\bibinfo{author}{\bibfnamefont{P.}~\bibnamefont{Abratenko}}
  \bibnamefont{et~al.} (\bibinfo{collaboration}{MicroBooNE}),
  \bibinfo{journal}{Phys. Rev. Lett.} \textbf{\bibinfo{volume}{127}},
  \bibinfo{pages}{151803} (\bibinfo{year}{2021}), \eprint{2106.00568}.

\bibitem[{\citenamefont{Ariga et~al.}(2019)}]{FASER:2018eoc}
\bibinfo{author}{\bibfnamefont{A.}~\bibnamefont{Ariga}} \bibnamefont{et~al.}
  (\bibinfo{collaboration}{FASER}), \bibinfo{journal}{Phys. Rev. D}
  \textbf{\bibinfo{volume}{99}}, \bibinfo{pages}{095011}
  (\bibinfo{year}{2019}), \eprint{1811.12522}.

\bibitem[{\citenamefont{Gligorov et~al.}(2018)\citenamefont{Gligorov, Knapen,
  Papucci, and Robinson}}]{Gligorov:2017nwh}
\bibinfo{author}{\bibfnamefont{V.~V.} \bibnamefont{Gligorov}},
  \bibinfo{author}{\bibfnamefont{S.}~\bibnamefont{Knapen}},
  \bibinfo{author}{\bibfnamefont{M.}~\bibnamefont{Papucci}}, \bibnamefont{and}
  \bibinfo{author}{\bibfnamefont{D.~J.} \bibnamefont{Robinson}},
  \bibinfo{journal}{Phys. Rev. D} \textbf{\bibinfo{volume}{97}},
  \bibinfo{pages}{015023} (\bibinfo{year}{2018}), \eprint{1708.09395}.

\bibitem[{\citenamefont{Ahdida et~al.}(2021)}]{SHiP:2020vbd}
\bibinfo{author}{\bibfnamefont{C.}~\bibnamefont{Ahdida}} \bibnamefont{et~al.}
  (\bibinfo{collaboration}{SHiP}), \bibinfo{journal}{Eur. Phys. J. C}
  \textbf{\bibinfo{volume}{81}}, \bibinfo{pages}{451} (\bibinfo{year}{2021}),
  \eprint{2011.05115}.

\bibitem[{\citenamefont{Feng et~al.}(2018)\citenamefont{Feng, Galon, Kling, and
  Trojanowski}}]{Feng:2017vli}
\bibinfo{author}{\bibfnamefont{J.~L.} \bibnamefont{Feng}},
  \bibinfo{author}{\bibfnamefont{I.}~\bibnamefont{Galon}},
  \bibinfo{author}{\bibfnamefont{F.}~\bibnamefont{Kling}}, \bibnamefont{and}
  \bibinfo{author}{\bibfnamefont{S.}~\bibnamefont{Trojanowski}},
  \bibinfo{journal}{Phys. Rev. D} \textbf{\bibinfo{volume}{97}},
  \bibinfo{pages}{055034} (\bibinfo{year}{2018}), \eprint{1710.09387}.

\bibitem[{\citenamefont{Kling and Trojanowski}(2021)}]{Kling:2021fwx}
\bibinfo{author}{\bibfnamefont{F.}~\bibnamefont{Kling}} \bibnamefont{and}
  \bibinfo{author}{\bibfnamefont{S.}~\bibnamefont{Trojanowski}},
  \bibinfo{journal}{Phys. Rev. D} \textbf{\bibinfo{volume}{104}},
  \bibinfo{pages}{035012} (\bibinfo{year}{2021}), \eprint{2105.07077}.

\bibitem[{\citenamefont{Carena et~al.}(2023)\citenamefont{Carena, Kozaczuk,
  Liu, Ou, Ramsey-Musolf, Shelton, Wang, and Xie}}]{Carena:2022yvx}
\bibinfo{author}{\bibfnamefont{M.}~\bibnamefont{Carena}},
  \bibinfo{author}{\bibfnamefont{J.}~\bibnamefont{Kozaczuk}},
  \bibinfo{author}{\bibfnamefont{Z.}~\bibnamefont{Liu}},
  \bibinfo{author}{\bibfnamefont{T.}~\bibnamefont{Ou}},
  \bibinfo{author}{\bibfnamefont{M.~J.} \bibnamefont{Ramsey-Musolf}},
  \bibinfo{author}{\bibfnamefont{J.}~\bibnamefont{Shelton}},
  \bibinfo{author}{\bibfnamefont{Y.}~\bibnamefont{Wang}}, \bibnamefont{and}
  \bibinfo{author}{\bibfnamefont{K.-P.} \bibnamefont{Xie}},
  \bibinfo{journal}{LHEP} \textbf{\bibinfo{volume}{2023}}, \bibinfo{pages}{432}
  (\bibinfo{year}{2023}), \eprint{2203.08206}.

\bibitem[{\citenamefont{Kozaczuk et~al.}(2020)\citenamefont{Kozaczuk,
  Ramsey-Musolf, and Shelton}}]{Kozaczuk:2019pet}
\bibinfo{author}{\bibfnamefont{J.}~\bibnamefont{Kozaczuk}},
  \bibinfo{author}{\bibfnamefont{M.~J.} \bibnamefont{Ramsey-Musolf}},
  \bibnamefont{and} \bibinfo{author}{\bibfnamefont{J.}~\bibnamefont{Shelton}},
  \bibinfo{journal}{Phys. Rev. D} \textbf{\bibinfo{volume}{101}},
  \bibinfo{pages}{115035} (\bibinfo{year}{2020}), \eprint{1911.10210}.

\bibitem[{\citenamefont{Carena et~al.}(2020)\citenamefont{Carena, Liu, and
  Wang}}]{Carena:2019une}
\bibinfo{author}{\bibfnamefont{M.}~\bibnamefont{Carena}},
  \bibinfo{author}{\bibfnamefont{Z.}~\bibnamefont{Liu}}, \bibnamefont{and}
  \bibinfo{author}{\bibfnamefont{Y.}~\bibnamefont{Wang}},
  \bibinfo{journal}{JHEP} \textbf{\bibinfo{volume}{08}}, \bibinfo{pages}{107}
  (\bibinfo{year}{2020}), \eprint{1911.10206}.

\bibitem[{\citenamefont{Liu et~al.}(2022{\natexlab{a}})\citenamefont{Liu, Yang,
  and Sun}}]{Liu:2022nvk}
\bibinfo{author}{\bibfnamefont{W.}~\bibnamefont{Liu}},
  \bibinfo{author}{\bibfnamefont{A.}~\bibnamefont{Yang}}, \bibnamefont{and}
  \bibinfo{author}{\bibfnamefont{H.}~\bibnamefont{Sun}},
  \bibinfo{journal}{Phys. Rev. D} \textbf{\bibinfo{volume}{105}},
  \bibinfo{pages}{115040} (\bibinfo{year}{2022}{\natexlab{a}}),
  \eprint{2205.08205}.

\bibitem[{\citenamefont{Kanemura and Li}(2024)}]{Kanemura:2023jiw}
\bibinfo{author}{\bibfnamefont{S.}~\bibnamefont{Kanemura}} \bibnamefont{and}
  \bibinfo{author}{\bibfnamefont{S.-P.} \bibnamefont{Li}},
  \bibinfo{journal}{JCAP} \textbf{\bibinfo{volume}{03}}, \bibinfo{pages}{005}
  (\bibinfo{year}{2024}), \eprint{2308.16390}.

\bibitem[{\citenamefont{Baldes et~al.}(2021)\citenamefont{Baldes, Blasi,
  Mariotti, Sevrin, and Turbang}}]{Baldes:2021vyz}
\bibinfo{author}{\bibfnamefont{I.}~\bibnamefont{Baldes}},
  \bibinfo{author}{\bibfnamefont{S.}~\bibnamefont{Blasi}},
  \bibinfo{author}{\bibfnamefont{A.}~\bibnamefont{Mariotti}},
  \bibinfo{author}{\bibfnamefont{A.}~\bibnamefont{Sevrin}}, \bibnamefont{and}
  \bibinfo{author}{\bibfnamefont{K.}~\bibnamefont{Turbang}},
  \bibinfo{journal}{Phys. Rev. D} \textbf{\bibinfo{volume}{104}},
  \bibinfo{pages}{115029} (\bibinfo{year}{2021}), \eprint{2106.15602}.

\bibitem[{\citenamefont{Azatov et~al.}(2021{\natexlab{a}})\citenamefont{Azatov,
  Vanvlasselaer, and Yin}}]{Azatov:2021irb}
\bibinfo{author}{\bibfnamefont{A.}~\bibnamefont{Azatov}},
  \bibinfo{author}{\bibfnamefont{M.}~\bibnamefont{Vanvlasselaer}},
  \bibnamefont{and} \bibinfo{author}{\bibfnamefont{W.}~\bibnamefont{Yin}},
  \bibinfo{journal}{JHEP} \textbf{\bibinfo{volume}{10}}, \bibinfo{pages}{043}
  (\bibinfo{year}{2021}{\natexlab{a}}), \eprint{2106.14913}.

\bibitem[{\citenamefont{Azatov et~al.}(2021{\natexlab{b}})\citenamefont{Azatov,
  Vanvlasselaer, and Yin}}]{Azatov:2021ifm}
\bibinfo{author}{\bibfnamefont{A.}~\bibnamefont{Azatov}},
  \bibinfo{author}{\bibfnamefont{M.}~\bibnamefont{Vanvlasselaer}},
  \bibnamefont{and} \bibinfo{author}{\bibfnamefont{W.}~\bibnamefont{Yin}},
  \bibinfo{journal}{JHEP} \textbf{\bibinfo{volume}{03}}, \bibinfo{pages}{288}
  (\bibinfo{year}{2021}{\natexlab{b}}), \eprint{2101.05721}.

\bibitem[{\citenamefont{Baldes et~al.}(2023)\citenamefont{Baldes, Dichtl,
  Gouttenoire, and Sala}}]{Baldes:2023fsp}
\bibinfo{author}{\bibfnamefont{I.}~\bibnamefont{Baldes}},
  \bibinfo{author}{\bibfnamefont{M.}~\bibnamefont{Dichtl}},
  \bibinfo{author}{\bibfnamefont{Y.}~\bibnamefont{Gouttenoire}},
  \bibnamefont{and} \bibinfo{author}{\bibfnamefont{F.}~\bibnamefont{Sala}}
  (\bibinfo{year}{2023}), \eprint{2306.15555}.

\bibitem[{\citenamefont{Ai et~al.}(2024)\citenamefont{Ai, Fairbairn, Mimasu,
  and You}}]{Ai:2024ikj}
\bibinfo{author}{\bibfnamefont{W.-Y.} \bibnamefont{Ai}},
  \bibinfo{author}{\bibfnamefont{M.}~\bibnamefont{Fairbairn}},
  \bibinfo{author}{\bibfnamefont{K.}~\bibnamefont{Mimasu}}, \bibnamefont{and}
  \bibinfo{author}{\bibfnamefont{T.}~\bibnamefont{You}} (\bibinfo{year}{2024}),
  \eprint{2406.20051}.

\bibitem[{\citenamefont{Liu et~al.}(2022{\natexlab{b}})\citenamefont{Liu, Bian,
  Cai, Guo, and Wang}}]{Liu:2021svg}
\bibinfo{author}{\bibfnamefont{J.}~\bibnamefont{Liu}},
  \bibinfo{author}{\bibfnamefont{L.}~\bibnamefont{Bian}},
  \bibinfo{author}{\bibfnamefont{R.-G.} \bibnamefont{Cai}},
  \bibinfo{author}{\bibfnamefont{Z.-K.} \bibnamefont{Guo}}, \bibnamefont{and}
  \bibinfo{author}{\bibfnamefont{S.-J.} \bibnamefont{Wang}},
  \bibinfo{journal}{Phys. Rev. D} \textbf{\bibinfo{volume}{105}},
  \bibinfo{pages}{L021303} (\bibinfo{year}{2022}{\natexlab{b}}),
  \eprint{2106.05637}.

\bibitem[{\citenamefont{Lewicki et~al.}(2024)\citenamefont{Lewicki, Toczek, and
  Vaskonen}}]{Lewicki:2024ghw}
\bibinfo{author}{\bibfnamefont{M.}~\bibnamefont{Lewicki}},
  \bibinfo{author}{\bibfnamefont{P.}~\bibnamefont{Toczek}}, \bibnamefont{and}
  \bibinfo{author}{\bibfnamefont{V.}~\bibnamefont{Vaskonen}}
  (\bibinfo{year}{2024}), \eprint{2402.04158}.

\bibitem[{\citenamefont{Cai et~al.}(2024)\citenamefont{Cai, Hao, and
  Wang}}]{Cai:2024nln}
\bibinfo{author}{\bibfnamefont{R.-G.} \bibnamefont{Cai}},
  \bibinfo{author}{\bibfnamefont{Y.-S.} \bibnamefont{Hao}}, \bibnamefont{and}
  \bibinfo{author}{\bibfnamefont{S.-J.} \bibnamefont{Wang}},
  \bibinfo{journal}{Sci. China Phys. Mech. Astron.}
  \textbf{\bibinfo{volume}{67}}, \bibinfo{pages}{290411}
  (\bibinfo{year}{2024}), \eprint{2404.06506}.

\bibitem[{\citenamefont{Xie and Zhan}(2026)}]{Xie:2026vor}
\bibinfo{author}{\bibfnamefont{K.-P.} \bibnamefont{Xie}} \bibnamefont{and}
  \bibinfo{author}{\bibfnamefont{C.-H.} \bibnamefont{Zhan}}
  (\bibinfo{year}{2026}), \eprint{2603.09126}.

\bibitem[{\citenamefont{Carrington}(1992)}]{Carrington:1991hz}
\bibinfo{author}{\bibfnamefont{M.~E.} \bibnamefont{Carrington}},
  \bibinfo{journal}{Phys. Rev. D} \textbf{\bibinfo{volume}{45}},
  \bibinfo{pages}{2933} (\bibinfo{year}{1992}).

\end{thebibliography}

\end{document}